\documentclass[draft]{agujournal2019}
\usepackage{url} 
\usepackage{lineno}
\usepackage[inline]{trackchanges} 
\usepackage{soul}
\usepackage{amsmath}

\usepackage{lmodern}
\usepackage{float}

\linenumbers

\def\wn{wavenumber }
\def\wns{wavenumbers }

\def\eql{equatorial }

\def\st{stratosphere }

\def\Deg{\textdegree}
\def\bkg{background }

\draftfalse

\journalname{Journal of Advances in Modeling Earth Systems (JAMES)}

\nolinenumbers

\begin{document}

%
%


\title{Wave Decomposition and Power-Law Scaling of Vertical Momentum Fluxes on the Sphere: Application to the QBO}

%
%




\authors{Valentino Neduhal\affil{1}, Nedjeljka \v{Z}agar\affil{1} and Frank Lunkeit\affil{1}}


\affiliation{1}{Department of Earth System Sciences, University of Hamburg, Germany}




\correspondingauthor{Nedjeljka \v{Z}agar}{nedjeljka.zagar@uni-hamburg.de}



\begin{keypoints}
\item A new method computes vertical momentum flux (VMF) on the sphere using horizontal and vertical velocities of wave modes. 
\item At each zonal wavenumber, VMF cospectra are proportional to wave frequency and  mechanical energy of the horizontal flow. 
\item Most lower stratospheric VMF within 5\Deg S–5\Deg N in ECMWF analyses is carried by Kelvin and inertia-gravity waves with scales $>$ 2000 km.
\end{keypoints}

%
%

%
%

\begin{abstract}
Vertical momentum flux (VMF) by atmospheric waves has been estimated using scale cutoff assumptions in wavenumber space or wavenumber-frequency space, with each wavenumber-frequency point attributed to a single wave mode. We introduce a unified spherical framework that decomposes horizontal and vertical wave velocities into Rossby (G), eastward- and westward-propagating inertia-gravity (IG), Kelvin, and mixed Rossby gravity (MRG) waves as functions of latitude and altitude. The new framework represents a departure from monochromatic view of VMF by allowing different wave types to coexist across scales, which is especially relevant in the tropics. We show that the VMF cospectrum ($F_k$) in the hydrostatic atmosphere is proportional to the horizontal flow energy ($I_k$) and wave frequency ($\nu$): $F_k \propto \nu_k I_k$, leading to wave-dependent VMF power-law scalings with the zonal wavenumber $k$.  Application to ECMWF analysis data reveals altitude, latitude, seasonal, and QBO phase variations of VMF associated with the coexistence of different wave types at large scales. The IG modes dominate VMF in the tropical lower stratosphere at subsynoptic scales, whereas IG and Kelvin waves together account for most VMF at larger scales. MRG wave VMF is more than an order of magnitude smaller than Kelvin wave VMF during both QBO phases. The new VMF decomposition provides a physically based framework for evaluating resolved and parameterized wave contributions without predefined scale or frequency cutoffs.
\end{abstract} %

\section*{Plain Language Summary}
The vertical transport of horizontal momentum (vertical momentum flux; VMF) by atmospheric waves is one of the key drivers of stratospheric circulation. Assessing VMFs and attributing them to wave motions of different scales and frequencies involves numerous simplifications. Typical assumptions are that motions with scales smaller than 2000~km in the extratropics involve only inertia-gravity (IG) or gravity waves, whereas only Rossby waves reside at larger scales. In the tropics, flows within predefined frequency and spatial-scale ranges are assigned to a single wave type among Rossby, IG, Kelvin, and mixed Rossby-gravity (MRG) waves. The effects of these cutoff scales on estimated VMFs and their uncertainties in climate models that rely on VMF parameterizations remain poorly understood. Here, we introduce a method for VMF computation that does not rely on these cutoffs. Instead, our method uses a spherical framework to decompose atmospheric flows into specific wave types that coexist across different scales, which is particularly important in the tropics. Furthermore, the method yields wave-specific power-law behavior based on the proportionality between VMF cospectra, horizontal flow energy, and wave frequency. We applied the new method to two-year ECMWF analyses to examine the theoretical power-law scalings and the role of the cutoff scale.
 
%
%

%


%
%
%
%


\section{Introduction}

The transport of the horizontal momentum by vertical velocity of the waves --  the vertical momentum flux (VMF) -- remains a challenge to measure and 
simulate by numerical models \cite<e.g.,>[]{geller2013comparison,holt2022evaluation}. 
The transport is carried out by waves of different scales and propagation properties: small-scale gravity waves unaffected by rotation, inertia-gravity (IG) waves, Rossby waves and equatorially trapped Kelvin waves and mixed Rossby–gravity (MRG) waves. 
The VMFs originating from the troposphere are believed to be the main wave forcing of the middle atmosphere \cite<e.g.,>[]{kawatani2010roles,McLandressetal2013,KimChun2015}. 
The key question regarding vertically propagating waves and associated VMFs in weather and climate models is: what fraction of VMF is carried by gravity and IG waves that require parameterization, compared with resolved waves of different types? Addressing this question requires a scale- and wave-mode-dependent decomposition of atmospheric circulation into Rossby and IG waves in the extratropics, and into Rossby, IG, MRG, and Kelvin waves in the tropics.

Traditionally, tropical and extratropical VMFs have been investigated using different methodologies. Here, we present a unified approach for decomposing VMFs on the sphere that permits the coexistence of the Rossby, MRG, IG, and Kelvin waves across a broad range of spatial scales found in data \cite<e.g.,>[]{Hayashi1981,WheelerKiladis99,Tindalletal2006b,Yangetal2007a,HendonWheeler2008,Zagaretal2009b}. Our approach accounts for the decomposition of both the horizontal and vertical velocity fields within a hydrostatic framework, which is applicable even to current kilometer-scale atmospheric models \cite<e.g.,>[]{Learetal2024,Vivodaetal2026}. 
In the remainder of this introduction, we motivate our approach by discussing recent progress  in wave decomposition and VMF diagnostics, scale distributions of vertical and horizontal kinetic energy across for the different wave types, and by highlighting outstanding scientific questions. 

In theory, the horizontal wind perturbation $(u',v')$ associated with gravity or IG waves at a single time step is assumed to be a departure between the full wind $(u,v)$ and a slowly varying or steady large-scale background state $(\bar{u},\bar{v})$, $u=\bar{u}+u'$ and $v=\bar{v}+v'$. In the hydrostatic atmosphere, the vertical velocity $w$ has the zero background, $\bar{w}=0$. For reanalysis or model data, $(u',v')$ have commonly been estimated using a cutoff scale of the zonal wavenumber using the Fourier series expansion along latitude circles or the total (or global) \wn associated with the spherical harmonics expansion. A typical cutoff scale is about 2000~km which corresponds to the zonal wavenumber 20 at the equator \cite<e.g.,>[]{Polichtchouketal2022}, although  the applied scale varies from the zonal \wns 6 \cite<e.g.,>[]{Preusseetal2014} and 15 \cite<e.g.,>[]{Jewtoukoffetal2015} to the horizontal \wn 40 \cite{wei2022global}.

\begin{figure}[h]
\centering
\includegraphics[width=0.99\textwidth]{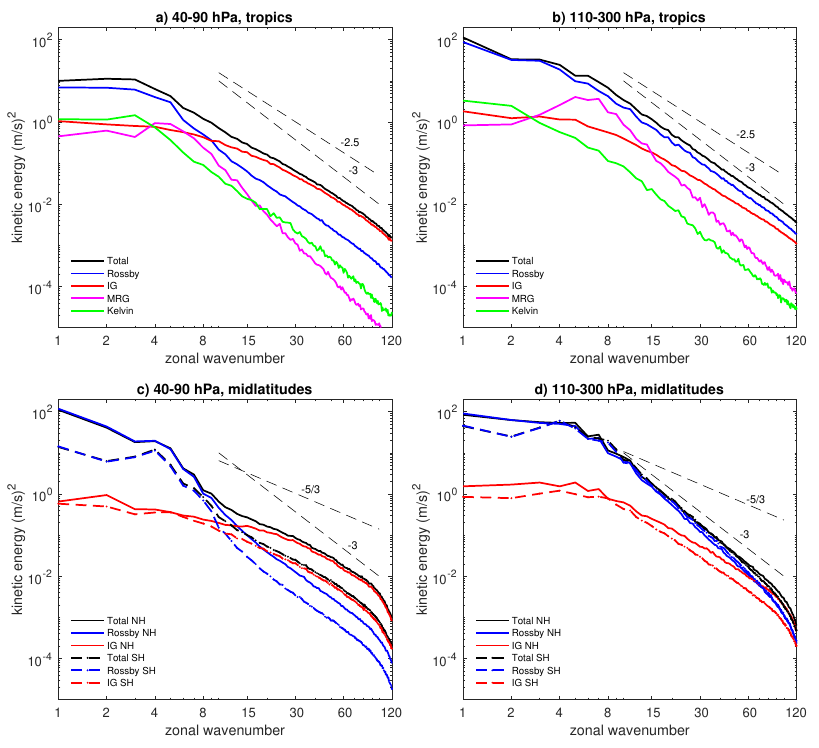}
\caption{Horizontal kinetic energy (HKE) spectra as a function of zonal wavenumber, based on ECMWF 12~UTC analyses in January 2021. The spectra for the Rossby, inertia-gravity (IG), Kelvin, and mixed Rossby-gravity (MRG) modes are obtained using the spherical wavenumber decomposition, explained in Method section. Tropical and midlatitude spectra are averaged over latitudes 20\Deg N--20\Deg S and 30\Deg--70\Deg in the Northern Hemisphere (NH) and Southern Hemisphere (SH), respectively. "Total" refers to the total HKE, which is equal to the sum of the Rossby, IG, MRG, and Kelvin components.}
\label{fig1}
\end{figure}

The application of the cutoff scale relies on separation of fast, divergence-dominated dynamics and slow, advective or vorticity-dominated motion on the $f$-plane or midlatitude $\beta$-plane. Even though the separation between fast and slow dynamics is exact only in the linearized system on $f$-plane \cite<e.g.,>[]{Stephanetal2021}, the application of this concept has been useful and illuminative for many problems in atmospheric dynamics in extratropics. The most canonical example is the dynamics described by the conservation of potential vorticity \cite<e.g.,>[]{Hoskins1985}. More recently, \citeA{Zagaretal2017} suggested using the crossing scale between the Rossby and IG wave modes, that is, the scale at which the energy associated with vorticity becomes smaller than the energy associated with divergent motions, as the cutoff scale. This scale varies with altitude and flow characteristics and generally shifts towards larger scales from the upper troposphere to the stratosphere \cite<e.g.,>[]{Blazicaetal2013,Burgess2013}, as illustrated in Figure~\ref{fig1}. \citeA{wei2022global} and \citeA{gupta2024estimates} accounted for a continuous spectrum of vortical and divergent flows at subsynoptic scales by applying the cutoff scale to the divergent wind following the Helmholtz decomposition.

The cutoff scale cannot be applied in the tropics. Small-scale gravity waves can still be described using the same equations as those applicable in the midlatitudes, with $f=0$ \cite<e.g.,>[]{Nappo}, but the scale separation between IG and Rossby waves at larger scales is lost, while Kelvin and MRG waves fill the frequency gap between the Rossby and IG waves \cite<e.g.,>[]{Matsuno66,Boyd2018}. A canonical example is the Kelvin wave, whose vorticity–divergence ratio varies strongly with zonal scale \cite{Neduhaletal2024}. A substantial fraction of its variance spectrum is quasi-stationary \cite{Zagaretal2022,Holubeetal2025}. The MRG wave is largely a vorticity wave at all scales \cite{Neduhaletal2024}, but with much greater phase speeds compared to the \eql Rossby waves and largely different variance spectra in the upper troposphere and stratosphere \cite{Mahoetal2025}. The two waves are important contributors to the quasi-biennial oscillation \cite<QBO;>[]{baldwin2001quasi} which remains a challenge for climate models \cite<e.g.,>[]{richter2020progress,butchart2020qbo}. The relative roles of the horizontal wave scale and wave types for driving the QBO remains relatively unclear and model and method dependent \cite{sato1997estimates,ern2009,ern2009quantification,kawatani2010roles,KimChun2015,kim2015momentum,anstey2016simulating,holt2022evaluation,pahlavan2021revisitingPART2}. 
For example, \citeA{polichtchouk2023resolved} showed that increasing the horizontal resolution from 9~km to about 1~km in the ECMWF IFS model does not change the fluxes in the tropical stratosphere highlighting the combined effect from small-scale and large- and mesoscale waves in driving tropical middle atmosphere. 

More generally, atmospheric wave motions in analyses and prediction models span a continuous range of scales, with wave amplitudes dependent on latitude and background flow. This is illustrated in Figure~\ref{fig1}, which shows the horizontal kinetic energy (HKE) spectra for different waves in the tropical and extratropical upper troposphere and lower stratosphere (UTLS). For example, MRG waves in the upper tropical troposphere (Figure~\ref{fig1}b) show a pronounced HKE peak at synoptic scales that is absent in the stratosphere (Figure~\ref{fig1}a) and has been explained by scale selection associated with wave-mean flow interactions \cite{Mahoetal2025}. Large amplitude, planetary-scale Rossby wave variability in the tropical upper troposphere in Figure~\ref{fig1}b is associated with the boreal winter westerlies over the central and eastern Pacific in January 2021 (Figure S1 in the Supplementary information; SI). In the midlatitude stratosphere, the winter hemisphere contains stronger subsynoptic variability in IG modes at all scales compared with the summer hemisphere and the upper troposphere (Figure~\ref{fig1}c-d). This is associated with the ageostrophic component of large-scale dynamics and IG waves at subsynoptic scales.   

\begin{figure}[h]
\centering
\includegraphics[width=0.99\textwidth]{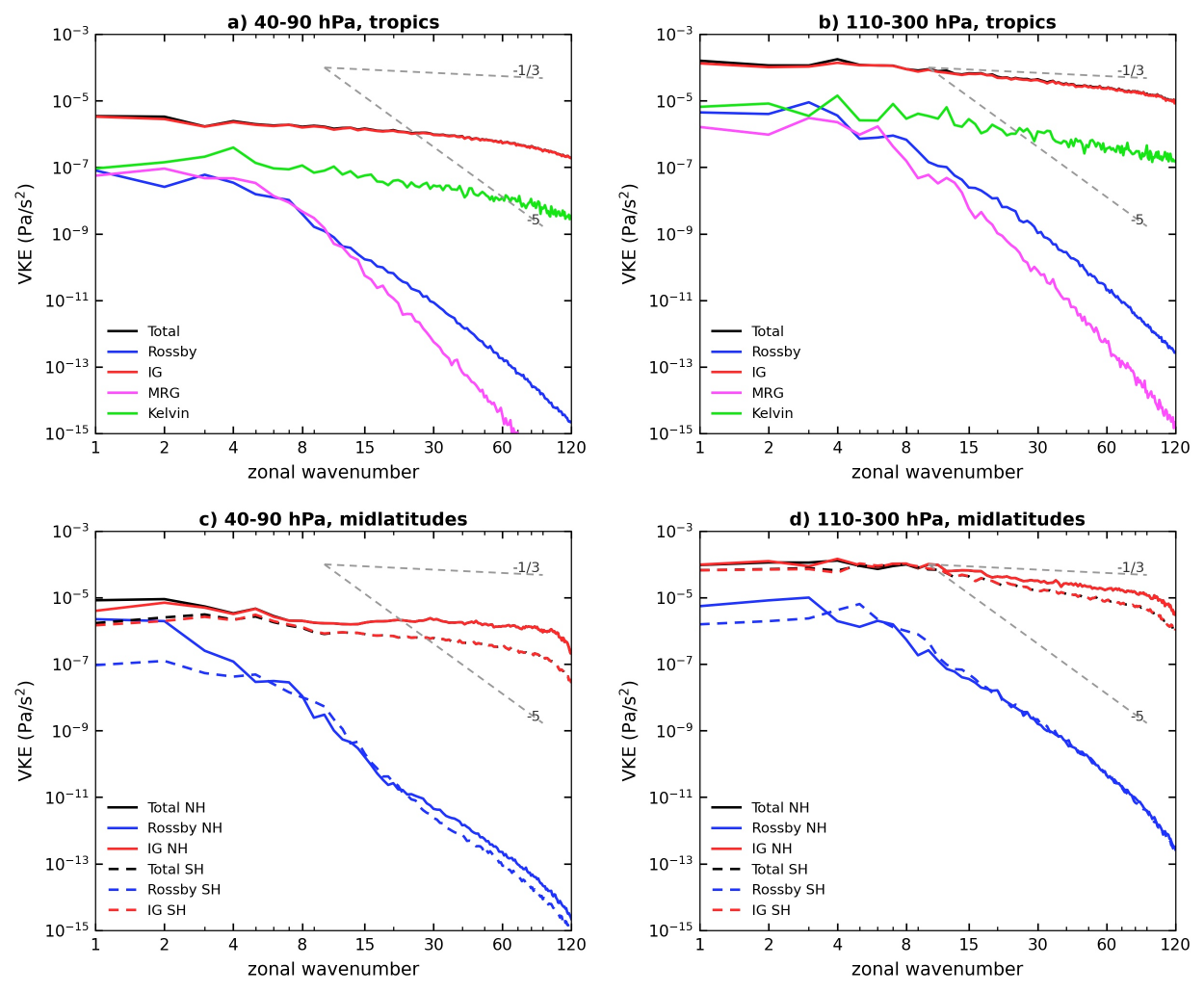}
\caption{As in Figure~1, but the vertical kinetic energy (VKE) for the pressure vertical velocity $\omega$.}
\label{fig2}
\end{figure}

Spectra of the kinetic energy associated with vertical motions (i.e., vertical kinetic energy, VKE) for the same layers and latitudes are shown in Figure~\ref{fig2}. The largest VKE occurs in the tropical upper troposphere, with a slightly red spectrum across the subsynoptic scales, consistent with a steeper-than-theoretically expected horizontal kinetic energy (HKE) spectrum in Figure~\ref{fig1} and the relationship $VKE \propto \nu^2 HKE$, where $\nu$ is the wave frequency \cite{Zagaretal2023}. Figure~\ref{fig2} shows that the VKE associated with Kelvin waves is smaller than that associated with IG modes, whereas the tropical MRG and Rossby wave VKEs have similarly small magnitudes. The midlatitude spectra reveal a gap between the winter- and summer-hemisphere IG VKE spectra at subsynoptic scales and Rossby VKE spectra at planetary scales. The hemispheric differences are more pronounced in the stratosphere (Figure~\ref{fig2}c–d). The presence of different wave types across scales implies VMF terms of the form $\overline{u'_R w'_{\mathrm{IG}}}$ and $\overline{v'_R w'_{\mathrm{IG}}}$. Such mixed terms have not been considered previously because of the use of a cutoff scale, below which all motions are assumed to belong to the IG or gravity-wave component in the extratropics.

Estimating VMF in the tropics requires large-scale equatorial wave filtering, which is not possible using the Helmholtz decomposition. Tropical VMFs have been estimated using wavenumber-frequency filtering \cite{Hayashi1981}, based on  Fourier series decomposition in longitude and time, combined with linear equatorial wave theory  \cite{WheelerKiladis99,kiladis2009convectively}. 
In this approach, subsets of the zonal wavenumber–frequency space are assigned to individual wave types \cite{KimChun2015,holt2022evaluation,pahlavan2021revisitingPART2}; that is, wave forcing within a certain range of zonal \wns and wave periods is assumed to belong to a single wave type. For example, \citeA{pahlavan2021revisitingPART2} define the zonal \wns $1 \le k \le 20$ with periods longer than 2.5 days, eastward propagation and \eql symmetry as Kelvin waves, whereas waves with $|k|\le 20$, periods of 2-10 days, and equatorial antisymmetry are associated with MRG waves. The remaining power in the \wn range $|k| \le 20$ is considered to be associated with Rossby waves if periods are longer than 2.5 days and with IG waves for periods shorter than 2.5 days. All perturbations with $|k| > 20$, regardless of frequency, are classified as small-scale gravity waves. 
If the data contain antisymmetric Rossby waves  (even meridional modes, $n=2,4,..$,) or antisymmetric westward-propagating IG (WIG) waves (odd meridional modes, $n=1,3,..$) with periods below 10 days, this approach interprets them as MRG waves, while large-scale symmetric, eastward-propagating IG (EIG) waves (even meridional modes, $n=2,4,..$) with  periods longer than 2.5 days are interpreted as Kelvin waves. 
The magnitude and relevance of possible misattribution of VMF resulting from various assumptions is difficult to estimate. In addition to the lack of observations, especially wind observations, the assignment of horizontal momentum  to individual wave types depends on the \eql wave filtering approach \cite{knippertz2022intricacies}. 
An intercomparison study addressing these uncertainties has yet to be conducted. Uncertainties extend to the assumptions about vertical velocity, with various studies using either the full vertical velocity when evaluating the VMF terms  $\overline{u'w'}$  and $\overline{v'w'}$   \cite{Weietal2025} or applying the same cutoff scale used for the horizontal winds  \cite{Polichtchouketal2022,polichtchouk2023resolved}. 

This study eliminates several of these assumptions by considering both horizontal and vertical wave motions without scale restrictions. The paper is structured as follows. Section 2 describes the methodology and datasets, including the new theoretical framework for the VMF decomposition and the power-law scaling of VMF cospectra. We demonstrate the capabilities of the new approach by examining the global and zonal-mean fluxes in ECMWF analyses and highlighting their physical interpretability (Section 3). Our initial applications of the new method focus on the following questions: (i) How do the spectral properties of the diagnosed VFMs compare with the scaling laws predicted by our methodology (Section 4)? (ii) Which wave types are associated with VMFs during the easterly and westerly QBO phases in 2019-2021 (Section 5)? (iii) What effect does a cutoff scale have on the quantitative assignment of VMFs to the different wave types (Section 6)? Finally, Section 7 provides a summary, conclusions, and an outlook.

\section{Methods}\label{Sec:paper2-methods}

The derivation of the 3D spherical VMF starts from the global primitive equations on the sphere, linearized around the state of rest. Their eigensolutions represent the five fundamental wave modes on the sphere: eastward- and westward-propagating IG waves (EIG and WIG, respectively), Rossby, MRG, and Kelvin waves. For a given frequency, their horizontal structures depend on the three spatial wave indices: zonal wavenumber $k$, meridional mode index $n$, and vertical mode index $m$. The derivation of these eigensolutions, also known as normal modes, builds on the early work of \citeA{Hough1898} and \citeA{Taylor1936} and subsequent developments such as \citeA{Kasahara76}, and is described in several textbooks \cite<e.g.,>[]{Pedlosky79,Daley91}. Here, we follow the derivation and notation of \citeA{Zagaretal2023}.

\subsection{Linear wave theory approach for the VMF decomposition}
\subsubsection{Velocity decomposition in pressure coordinates }\label{methods_1} 

The horizontal wind perturbations on the sphere can be seen as a superposition of linear waves, represented by Hough harmonics. They are equivalents of the cosine and sine functions along a latitude circle and the parabolic cylinder function decomposition on the meridionally infinite \eql $\beta$ plane \cite{Boyd2018}. In this representation, the two wind components are expanded as 
\begin{equation}
\label{expansionUV}
\left[
 \begin{array}{cc}
    u\left( \lambda,\varphi,p \right)\\
    v\left( \lambda,\varphi,p \right) 
  \end{array}
\right] = \sum_{k=-K}^{K} \sum_{n=1}^{R} \sum_{m=1}^{M} \, \sqrt{g\,D_m} \, \chi_{n}^{k}(m) \, \left[
\begin{array}{cc}
    U_n^k   \\
    \mathrm{i} V_n^k   
  \end{array}\right] (\varphi;m) \, G_m(p)  \,.
  \mathrm{e}^{\mathrm{i} k\lambda} \,,
\end{equation}
The orthogonal set of vertical structure functions is denoted by $G_m(p)$, where $m$ denotes the index of the $m$-th eigenvalue $D_m$, known as the equivalent depth, and $G_m(p)$ is the associated eigenfunction of the vertical structure equation \cite<e.g.,>[]{Staniforthetal1985}. The scaling factor $\sqrt{g\,D_m}$ removes the dimensions from the input data after the vertical projection, with $g$ denoting gravity. There are $M$ vertical structure functions and $2K$ waves, in addition to the zonal mean state $k=0$. 
The horizontal structure functions, i.e. Hough harmonics, are products of  the longitudinal waves and Hough functions which are defined for the zonal and meridional wind components by $U_n^k (\varphi;m)$ and $V_n^k (\varphi;m)$ and for the geopotential $Z_n^k (\varphi;m)$. The imaginary unit $\mathrm{i}=\sqrt{-1})$ in front of $V_{n}^k$ accounts for its $\pi/2$ shift with respect to $U_{n}^k$. The three Hough functions are normalized according to the L2 norm. In other words, wind and pressure (or temperature) perturbations associated with the  waves are coupled through a common expansion coefficient $\chi_{n}^{k}(m)$. 

The five wave types are included within index $n$. The maximum number of meridional modes in the expansion~(\ref{expansionUV}), $R$, is the sum of $N_R$ Rossby modes, including the mixed Rossby--gravity mode as the lowest meridional mode ($n=0$), $N_{EIG}$ EIG modes, including the Kelvin wave as the lowest meridional mode ($n=0$), and $N_{WIG}$ WIG modes; $R = N_R + N_{EIG} + N_{WIG}$. The notation follows the formulation in the MODES software described in \citeA{Zagaretal2015b} where a detailed discussion of the steps involved in the computation of $\chi_{n}^{k}(m)$ is provided. 

The modal framework has recently been extended to include horizontal wind divergence \cite{Neduhaletal2024} and  pressure vertical velocity $\omega$ \cite{Zagaretal2023}, the latter of which was used to produce Figure~\ref{fig2}. The elegance of the VKE decomposition in pressure coordinates motivates its use for the VMF decomposition as well, rather than height coordinate. 
According to the theory, $\omega$ is obtained from the complex expansion coefficients using the continuity equation:  
\begin{equation}
\label{expansionO}
\omega(\lambda,\varphi,p)=  - \sum_{k=-K}^{K}  \sum_{n=1}^{R} \sum_{m=1}^{M} \mathrm{i}\,\nu_{n}^{k}(m)  \, \chi_{n}^{k}(m) \, Z_{n}^{k}\left(\varphi;m\right) \, \mathrm{e}^{\mathrm{i} k\lambda} \, \int_{0}^{p} G_m(p') \, \mathrm{ d} p'  \,,
 \end{equation}
where  $\nu_{n}^{k}(m)$ is a frequency of wave mode $(k,n,m)$.

On the other hand, $u$, $v$ and $\omega$ along the latitude circle $\varphi$ can be expanded using the Fourier series as 
\begin{equation}
     \Bigg[
\begin{matrix}
\, u \, \\
\, v \, \\
\, \omega\,
\end{matrix} \Bigg](\lambda,\varphi,p) = \sum_{k=-K}^{K}  \,  
 \Bigg[
\begin{matrix}
\, \hat{u}_k \, \\
\, \hat{v}_k \, \\
\, \hat{\omega}_k\, 
\end{matrix} \Bigg] (\varphi,p) \, \mathrm{e}^{\mathrm{i} k\lambda} \,.
\label{1dfft}
\end{equation}
The comparison of (\ref{expansionUV}), (\ref{expansionO}) and (\ref{1dfft}) gives the following definitions of the spectral expansion coefficients $\hat{u}_{k}$ and $\hat{v}_{k}$: 
\begin{equation}
\Bigg[
\begin{matrix}
\hat{u}_k \\
\hat{v}_k 
\end{matrix} \Bigg](\varphi, p) = \sum_{m=1}^{M}  \, \sum_{n=1}^{R} \sqrt{g \,D_m}\,\, \chi_{n}^{k}(m)  \,  \, 
\label{eq4}
 \Bigg[
\begin{matrix}
U_n^k \\
\mathrm{i} V_n^k 
\end{matrix} \Bigg] (\varphi;m)\,  \,  G_{m}(p) \, ,
\end{equation}
while $\hat{\omega}_{k}$ is 
\begin{equation}
\label{eq3}
\hat{\omega}_k (\varphi,p)=   -  \mathrm{i} \sum_{m=1}^{M} \sum_{n=1}^{R}  \nu_{n}^{k}(m) \chi_{n}^{k}(m)  Z_{n}^{k}\left(\varphi;m\right) \int_{0}^{p} G_m(p') \mathrm{d} p'  \,.
 \end{equation} 
The zonal \wn components $u$, $v$ and $\omega$, defined by (\ref{eq4})-(\ref{eq3}), are additive, meaning that
\begin{subequations}
\label{eq:modes}
\begin{gather}
\hat{u}_k \left(\varphi,p \right) = \hat{u}_R +\hat{u}_K + \hat{u}_M + \hat{u}_E+\hat{u}_W  \,, \quad \text{and } \\
\hat{\omega}_k \left(\varphi,p \right) =  \hat{\omega}_R+ \hat{\omega}_K + \hat{\omega}_M + \hat{\omega}_E+\hat{\omega}_W \,, 
\end{gather}
\end{subequations}
where R, K, M, E, and W stand for Rossby, Kelvin, MRG, EIG and WIG modes, respectively. 
Each zonal \wn contains contributions from all meridional and vertical modes with the same $k$. The vertical mode summation represents the hydrostatic atmosphere as a set of shallow-water systems, each characterized by its own equivalent depth \cite{Taylor1936}. In this framework, vertical wave propagation is manifested by wave signals shifting between vertical modes, as demonstrated in 
\citeA{Zagaretal2009b} and \citeA{Zagaretal2017}. 

The horizontal kinetic energy per unit mass around a latitude circle, $E_H$, can be computed from (\ref{eq4}) as 
\begin{equation}
\label{hke-t}
E_H(\varphi,p) = \sum_{k=0}^K E_H^k (\varphi,p)=   \sum_{k=0}^K \left(1-\delta_{k0} \right) \left(|\hat{u}_k |^2 + |\hat{v}_k |^2\right) \,\,,
\end{equation}
where  $\delta_{k0}=1/2$ for $k=0$ and $0$ otherwise  (with $\hat{\omega_k}$ presented only for positive $k$). 
By limiting the summation over $n$ in Equation~(\ref{eq4}) to the MRG, Kelvin, IG or Rossby modes, $E_H$ associated with a single wave type is obtained, as presented in Figure~\ref{fig1}. 

Similarly, the vertical kinetic energy $E_V$ is computed as 
\begin{equation}
\label{vke}
E_V (\varphi,p) = \sum_{k=0}^K E_V^k =  \sum_{k=0}^K \left(1-\delta_{k0} \right) |\hat{\omega}_k |^2 \, ,
\end{equation}
as illustrated in Figure~\ref{fig2}.
A zonal wavenumber component of $E_H$ and $E_V$ is denoted by $E^k_H$ and $E^k_V$, respectively. 
For details on the derivation of $\omega$ and $E_V$ and their relation to the modal energy $I_k$, given by $I_k = \left(1-\delta_{k0} \right)\,|\chi_n^k(m)|^2$, see \citeA{Zagaretal2023}.

\subsubsection{Vertical momentum fluxes} 

The two components of the vertical flux of zonal and meridional momentum, $F_{u \omega}$ and $F_{v \omega}$, respectively, in the pressure coordinate system are defined as 
\begin{equation}
\label{F1}
 \left[
\begin{array}{cc}
    F_{u \omega}\\
    F_{v \omega}\\
  \end{array}\right]  = -\frac{1}{g}  \left[ 
\begin{array}{cc}
    \overline{u' \omega'}\\
    \overline{v' \omega'}\\
  \end{array}\right] \,,
 \end{equation} 
where $\overline{()}$ stands for averaging (zonal or temporal or both) and primes denote deviations from the zonal mean. 
The formulation (\ref{F1}) assumes $\omega \approx -\rho g w $ and gives fluxes in units of Pascals  \cite<e.g.,>[]{fritts2003gravity}. 

Using the Fourier series expansion (\ref{1dfft}) along the latitude circle for $u$, $v$ and $\omega$, and associated orthogonality conditions, Equation~(\ref{F1}) for the zonally-averaged VMF becomes 
\begin{equation}
\label{F2}
 \left[
\begin{array}{cc}
    F_{u \omega}\\
    F_{v \omega}\\
  \end{array}\right] (\varphi,p) =  -\frac{1}{g} \Re\Big( \sum_{k=1}^K \left[
\begin{array}{cc}
    \hat{u}_{k} \hat{\omega}_{k}^* \\
    \hat{v}_{k} \hat{\omega}_{k}^*  \\
  \end{array}\right] \Big) \,,
 \end{equation}
where $\Re$ denotes the real part and $*$ denotes the complex conjugate operation.

The right-hand side of Eq.~(\ref{F2}) satisfies the Parseval  theorem and corresponds to the VMF cospectrum, which is the real part of the VMF cross-spectrum. The cospectrum describes the in-phase part of the $u$, $v$ and $\omega$ signals contributing to the zonal mean VMF. Its components quantify scale contribution to the zonally averaged VMFs. 
A single component of the cross-spectrum, modal VMF, for the two horizontal wind components is given by 
\begin{equation}
\label{eq5}
\Bigg[
\begin{array}{cc}
     \hat{F}_{u \omega}\\
     \hat{F}_{v \omega}\\
  \end{array}\Bigg](k; \varphi, p)  =    \left[
\begin{array}{cc}
    \hat{u}_k \hat{\omega}^*_k \\
    \hat{v}_k \hat{\omega}^*_k \\
  \end{array}\right] \,.
\end{equation}
We refer to a single $k$ component of either the zonal of meridional VMF as $\hat{F}_k$. 
The two components of $\hat{F}_k$ are given by $\Re(\hat{F}_k)=\hat{u}_{k,r}\hat{\omega}_{k,r} + \hat{u}_{k,i}\hat{\omega}_{k,i}$ and $\Im(\hat{F}_k)=\hat{u}_{k,i}\hat{\omega}_{k,r} - \hat{u}_{k,r}\hat{\omega}_{k,i}$, where subscripts $r$ and $i$ denote the real and imaginary parts, respectively. While the real part of $\hat{F}_k$, $\Re(\hat{F}_k)$, defines the VMF cospectrum, its imaginary part, $\Im(\hat{F}_k)$, defines the VMF quadrature spectrum. It describes the out-of-phase components of the signals that do not contribute to the zonal-mean VMF. Note that Equations~(\ref{F1})-(\ref{eq5}) apply at every latitude, pressure level and wave type.  

Applying the decomposition (\ref{eq:modes}), the zonal and meridional components of the modal VMF defined by Equation~(\ref{eq5}) each include 25 elements; that is,  both the cospectrum and the quadrature spectrum can be decomposed into 25 terms for each horizontal wind components. The 25 components of $F_{u\omega}$ are listed in Table~\ref{T1}. 
We refer to terms with horizontal and vertical velocities from different wave types as "off-diagonal" or "mixed" terms,  and to terms where horizontal and vertical velocities are from the same wave type as "diagonal" terms.  
The decomposition of vertical velocity has not been considered in previous studies. 
If we use the same (full) $\omega$ for all wave types, the VMF decomposition contains only five elements.
\begin{table}[]
 \caption{Decomposition of the vertical flux of zonal momentum into the components based on the five wave types: K -- Kelvin, M -- mixed Rossby-gravity, E -- eastward-propagating inertia gravity, W -- eastward-propagating inertia gravity and R -- Rossby modes.}
\resizebox{\textwidth}{!}{%
\begin{tabular}{|c|c|c|c|c|c|}
\hline
$  $     &      $ \omega_{K}$   & $\omega_{M}$ & $\omega_{E}$ & $\omega_{W}$ & $\omega_{R}$ \\ \hline
$u_{K}$ & ${F}^{K-K} = \overline{u_K' \omega_{K}'}$ & $ {F}^{K-M} = \overline{u_K' \omega_{M}'}$ & $ {F}^{K-E} = \overline{u_K' \omega_{E}'}$ & $ {F}^{K-W} = \overline{u_K' \omega_{W}'}$ & $ {F}^{K-R} = \overline{u_K' \omega_{R}'}$ \\ \hline
$u_{M}$        &   $ {F}^{M-K} = \overline{u_{M}' \omega_{K}'}$ & $ {F}^{M-M} = \overline{u_{M}' \omega_{M}'}$ & $ {F}^{M-E} = \overline{u_{M}' \omega_{E}'}$ & $ {F}^{M-W} = \overline{u_{M}' \omega_{W}'}$ & $ {F}^{M-R} = \overline{u_{M}' \omega_{R}'}$ \\ \hline
$u_{E}$         & $ {F}^{E-K} = \overline{u_{E}' \omega_{K}'}$ & $ {F}^{E-M} = \overline{u_{E}' \omega_{M}'}$ & $ {F}^{E-E} = \overline{u_{E}' \omega_{E}'}$ & $ {F}^{E-W} = \overline{u_{E}' \omega_{W}'}$ & $ {F}^{E-R} = \overline{u_{E}' \omega_{R}'}$ \\ \hline
$u_{W}$          & $ {F}^{W-K} = \overline{u_{W}' \omega_{K}'}$ & $ {F}^{W-M} = \overline{u_{W}' \omega_{M}'}$ & $ {F}^{W-E} = \overline{u_{W}' \omega_{E}'}$ & $ {F}^{W-W} = \overline{u_{W}' \omega_{W}'}$ & $ {F}^{W-R} = \overline{u_{W}' \omega_{R}'}$ \\ \hline
$u_{R}$          & $ {F}^{R-K} = \overline{u_R' \omega_{K}'}$ & $ {F}^{R-M} = \overline{u_R' \omega_{M}'}$ & $ {F}^{R-E} = \overline{u_R' \omega_{E}'}$ & $ {F}^{R-W} = \overline{u_R' \omega_{W}'}$ & $ {F}^{R-R} = \overline{u_R' \omega_{R}'}$ \\ \hline
\end{tabular}}
\label{T1}
\end{table}

The diagonal terms of Table~\ref{T1} contain VMF contributions from meridional and vertical modes with the same $k$, as discussed above. 
For an individual hydrostatic wave mode, the $\pi/2$ phase shift between zonal and vertical velocities gives zero covariance $u'w'$. However, the covariance becomes nonzero when contributions from different vertical and meridional modes are summed. This differs from the common assumption about the total flux being a sum of fluxes carried by different monochromatic waves based on the polarization relations of the linearized non-hydrostatic, adiabatic, primitive equations on the $f$-plane \cite<e.g.,>[]{Nappo}. In this case, under the assumption that $m>1/H$ (where $H$ is the scale height), zonal and vertical velocities are completely in phase for an eastward and upward-propagating gravity wave, producing a nonzero covariance $\overline{u'w'}$ for a monochromatic wave. 
The hydrostatic approach is well suited for the reanalysis data such as ERA5, which are produced with the hydrostatic ECMWF IFS model. In the next section, we compare our hydrostatic estimates of VMFs with previous estimates of fluxes in the Boussinesq atmosphere \cite<e.g.,>[]{ern2009}.

\subsection{Cross spectra}

Different terms in Table~\ref{T1} can be investigated using the VMFs cospectra and quadrature spectra, which define spectral coherence and phase, respectively \cite{vonStorchZwiers2003}.  
Coherence analysis allows us to examine the spectrum-normalized contribution of VMF components at each wavenumber to the total covariance. Peaks of coherence indicate wavenumbers where substantial flux activity occurs. 

Squared spectral coherence (hereafter coherence) estimates the degree of linear association between the phases and amplitudes of horizontal and vertical velocities for each wavenumber.  The coherence can be computed using Equation~(\ref{eq5}) as:
\begin{equation}
\label{eqCOH}
CH(k; \varphi, p)  =  \frac{|\langle \hat{F}_k \rangle |^2}{\langle \hat{u}_k\hat{u}_k^* \rangle \langle \hat{\omega}_k\hat{\omega}_k^* \rangle} \,,
\end{equation}
where $\langle \cdot \rangle$ represents averaging in time and over the latitudinal band. 
Coherence values are always $0\leq CH \leq 1$, where $1$ implies that horizontal and vertical velocities are perfectly linearly related, while $0$ implies that they are linearly unrelated. 

However, the coherence of two linearly related signals that are in quadrature will also be $1$,
while their contribution to the cospectrum, and thus to the zonally averaged VMF, is zero. 
Therefore, we need the phase as an additional metric, which is defined as 
\begin{equation}
\label{eqPHS}
PH(k; p)  =  \bigg | \arctan \Big (-\frac{\langle \Im( \hat{F}_k) \rangle }{\langle \Re( \hat{F}_k) \rangle} \Big) \bigg | \,.
\end{equation}
$PH$ gives the relative phase shift between the horizontal and vertical velocities. For $0 < PH < \frac{\pi}{2}$ cospectrum has positive values that correspond to the negative contribution to the momentum fluxes because of the $-1/g$. For the same reason, $\frac{\pi}{2} < PH < \pi$ has negative values that correspond to the positive contribution to the momentum fluxes. The phase spectrum provides additional information only if signals are coherent.

\subsection{Power laws for the VMFs cospectra }

Following the definition (\ref{eq5}), we can define the zonal wavenumber $k$ component of the absolute VMF cospectrum as:
\begin{equation}
\label{slope1}
F_k = \sqrt{\left[ \Re(\hat{u}_k  \hat{\omega}^*_k) \right]^{2} + \left[ \Re(\hat{v}_k  \hat{\omega}^*_k) \right]^{2}},
\end{equation}
In Appendix A, we show that the absolute cospectrum is proportional to the square root of the horizontal and vertical kinetic energies: 
\begin{equation}
\label{slope2}
    F_k \propto \sqrt{E_H^k}\sqrt{E_V^k} \,,
\end{equation}
that are defined in Section \ref{methods_1}.   

In the hydrostatic atmosphere, the relationship between the VKE and HKE of a linear wave is determined by its frequency  \cite{Zagaretal2023}: 
\begin{equation}
\label{ene}
E^k_V \propto {| \nu_k \, \chi_k |}^2 \quad \text{i.e.} \quad {E_V}^k \propto  \nu_k^2 \, I_k \,\,,\quad \text{where} \quad I_k \propto {| \chi_k |}^2\,.
\end{equation}
The total modal energy $I_k$ is the sum of the horizontal kinetic energy and available potential energy (mechanical energy).  The replacement of ${| \nu_k \, \chi_k |}^2$ by $\nu_k^2 \, {| \chi_k |}^2$ in (\ref{ene}) means that the expression applies to the limit spectra \cite<>[Appendix B]{Zagaretal2023}. 

Using (\ref{ene}) in (\ref{slope2}) yields the universal power-law scaling for the VMF cospectrum:  
\begin{equation}
\label{slope3}
     F_k \propto \nu_k {|\chi_k |}^2 \quad \text{or}   \quad  F_k \propto \nu_k \, I_k \,,
\end{equation}
which states that the VMF cospectrum is proportional to the wave frequency and the mechanical energy (or, equivalently, the horizontal kinetic energy).

Using the frequency-dispersion relationships for various waves and their energy power laws yields power laws for the VMF cospectra, summarized in Table~\ref{T3}. 
Here we follow \citeA<>[]{Zagaretal2023} by using the theoretical power laws of $-3$ and $-5/3$ for Rossby and IG energy spectra in the inertial range, and by combining theoretical and observational constraints  for large scales in the tropics, based on the ECMWF analysis data (Figures~S3-S5 in SI). 
The frequency-dispersion relationships for different wave types and latitude belts, as well as their dependence on meridional and vertical mode, are discussed in \citeA{Zagaretal2023}. 

\begin{table}[]
 \caption{Power laws for the horizontal kinetic energy (HKE), frequency-wavenumber dependence ($\nu$-$k$ ) and VMF cospectra as a function of the zonal wavenumber.}
\resizebox{\textwidth}{!}{%
\begin{tabular}{c|c|c|c|c}
\hline
&   Wave            &   HKE      & $\nu$-$k$   & VMF cospectra \\ \hline
Inertial range   & Rossby            &  $I_R \propto k^{-3}$       & $\nu_R \propto k^{-1}$      & $F_k  \propto k^{-4}$\\
troposphere & IG  &  $I_{G} \propto k^{-5/3}$   & $\nu_{G} \propto k^{1}$  & $F_k \propto k^{-2/3}$\\   \hline
Tropical stratosphere,    & Rossby    &  $I_R \propto k^{-2}-k^{-3}$      & $\nu_R \propto k^{0}$      & $ F_k \propto k^{-2}-k^{-3}$\\
synoptic and  & IG &  $I_{G} \propto k^{-2}$ & $\nu_{G} \propto k^{1}$    & $F_k \propto k^{-1}$\\ 
subsynoptic scales & MRG      &  $I_{M} \propto k^{-3}$  & $\nu_{M} \propto k^{-1}$  & $F_k \propto k^{-4}$\\
& Kelvin   &  $I_{K} \propto k^{-2}-k^{-3}$    & $\nu_{K} \propto k^{1}$   & $F_k\propto k^{-1}-k^{-2}$\\ \hline
Tropical stratosphere,   & Rossby    &  $I_R \propto k^{-1}$      & $\nu_R \propto k^{0}$      & $ F_k \propto k^{-1}$\\ 
 planetary scales & IG  &  $I_{G} \propto k^{0}-k^{-1}$ & $\nu_{G} \propto k^{1/2}$    & $F_k \propto k^{-1/2}-k^{1/2}$\\
 (empirical estimate) & MRG   &  $I_{M} \propto k^{0}-k^{-1}$  & $\nu_{M} \propto k^{-1}$  & $F_k \propto k^{-2}-k^{-1}$\\
& Kelvin   &  $I_{K} \propto k^{-1}-k^{-2}$    & $\nu_{K} \propto k^{1}$   & $F_k \propto k^{-1}-k^{0}$\\ \hline
\end{tabular}}
\label{T3}
\end{table}

The structure of the HKE spectra varies with altitude, especially for MRG and Kelvin waves.  For example, there is a large difference between the kinetic energy spectra of MRG waves in the upper troposphere and upper stratosphere, associated with scale selection mechanism \cite<Figure 2 in >[]{Mahoetal2024}. 
The kinetic energy spectra at synoptic and subsynoptic scales in the upper tropical troposphere are similar for the two waves \cite<Figure 12 in >[]{Neduhaletal2024}, with slopes between -2 and -3. Similar spectra are found in the latest ECMWF data (see SI) and in the daily spectra available on the MODES webpage \url{https://modes.cen.uni-hamburg.de/products#ES}. In the tropical stratosphere, planetary-scale Kelvin and MRG waves are more energetic (per unit mass) \cite{Zagaretal2022,Mahoetal2024}, with spectra in the SI showing slopes ranging from -1 to -2 for the Kelvin wave and flatter spectra for MRG waves. However, the planetary-scale spectrum of \eql waves is generally uncertain because of uncertainties in tropical analyses \cite<e.g.,>[]{vzagar2025esa}, and an incomplete understanding of the dynamical mechanisms responsible for \eql wave structure and energy transfers. 

The power-law HKE scalings diagnosed from the ECMWF data and presented in the SI, when applied to Equation~(\ref{slope3}), imply $F_k \propto k^{-1}-k^0$ for planetary-scale Kelvin wave, whereas at synoptic and smaller scales  $F_k \propto k^{-2}-k^{-1}$. For the IG modes, $F_k \propto k^{-1/2}-k^{1/2}$ at planetary scales and  $F_k \propto k^{-1}$ at smaller scales. The MRG and Rossby waves follow $F_k \propto k^{-2}-k^{-1}$ and $F_k \propto k^{-1}$, respectively, at planetary scales and  $F_k \propto k^{-4}$ and $F_k \propto k^{-3}-k^{-2}$, respectively, at synoptic and smaller scales.

\subsection{Data and computation setup }

We use the ECMWF analyses produced with the IFS model version CY47R2 as part of a two-year Observing System Experiment performed to investigate the impact of assimilating Aeolus satellite wind measurements \cite{rennie2024}. The analyses are produced using the 12-hourly continuous 4D-Var \cite{Leanetal2020}, with an outer loop at resolution Tco639, which corresponds to a grid distance of approximately 18~km at the equator. The model has a "sponge layer" near the model top, implemented as fourth-order hyper-diffusion, with details described in \citeA{Neduhaletal2024}. More details about the analysis data can be found in \citeA{vzagar2025esa} and the references therein. 
The dataset includes both 00 and 12 UTC analyses from 1 July 2019 to 30 June 2021. The study period includes the QBO disruption in winter 2019/2020, which is evident in the results, but details are left for future studies. 

The input fields are defined on the 137 model levels, with the top level at 0.01 hPa. The zonal and meridional winds and geopotential are interpolated from the model's hybrid sigma-pressure levels to the pressure levels corresponding to the globally averaged pressure values at the full model levels. Horizontally, we use a significantly lower resolution, the N80  Gaussian grid, which consists of 320$\times$160 grid points and corresponds to a grid spacing of approximately 125~km at the equator. This resolution suffices for demonstrating the new method and for comparison with previous studies using reanalysis data. 

The projection to wave space includes $K=160$ zonal wavenumbers, $R=240$ meridional modes, consisting of 80 Rossby, 80 EIG and 80 WIG modes, and $M=50$  vertical modes. 
The truncation to $M=50$ vertical modes primarily affects the representation of the lower troposphere and  high latitudes, where the rapidly decreasing equivalent depth of higher vertical modes leads to equatorially trapped horizontal structures \cite{Zagaretal2009a}. However, the resulting differences in the middle atmosphere, which is the focus of our study, are small and can be neglected for the purposes of this study.
 
\section{Demonstration of the method using zonally averaged fluxes}

We now discuss the 25 components of the zonally averaged vertical flux of zonal momentum (VZMF), as defined in Table~\ref{T1}, and, analogously, the 25 components of the vertical flux of meridional momentum (VMMF), obtained using Equation~(\ref{F2}). The individual components are computed by applying Equations~(4--5) to the subsets of $n$ indices associated with each wave type, followed by zonal averaging according to Equation~(\ref{F2}). 
This section presents the results for a single date in 2020 to demonstrate the method and assess the physical interpretability of the results in comparison with previous studies. The following sections will present the evolution of the VZMF and VMMF over the two-year period and time-averaged properties. 

\begin{figure}[h]
\centering
\includegraphics[width=0.99\textwidth]{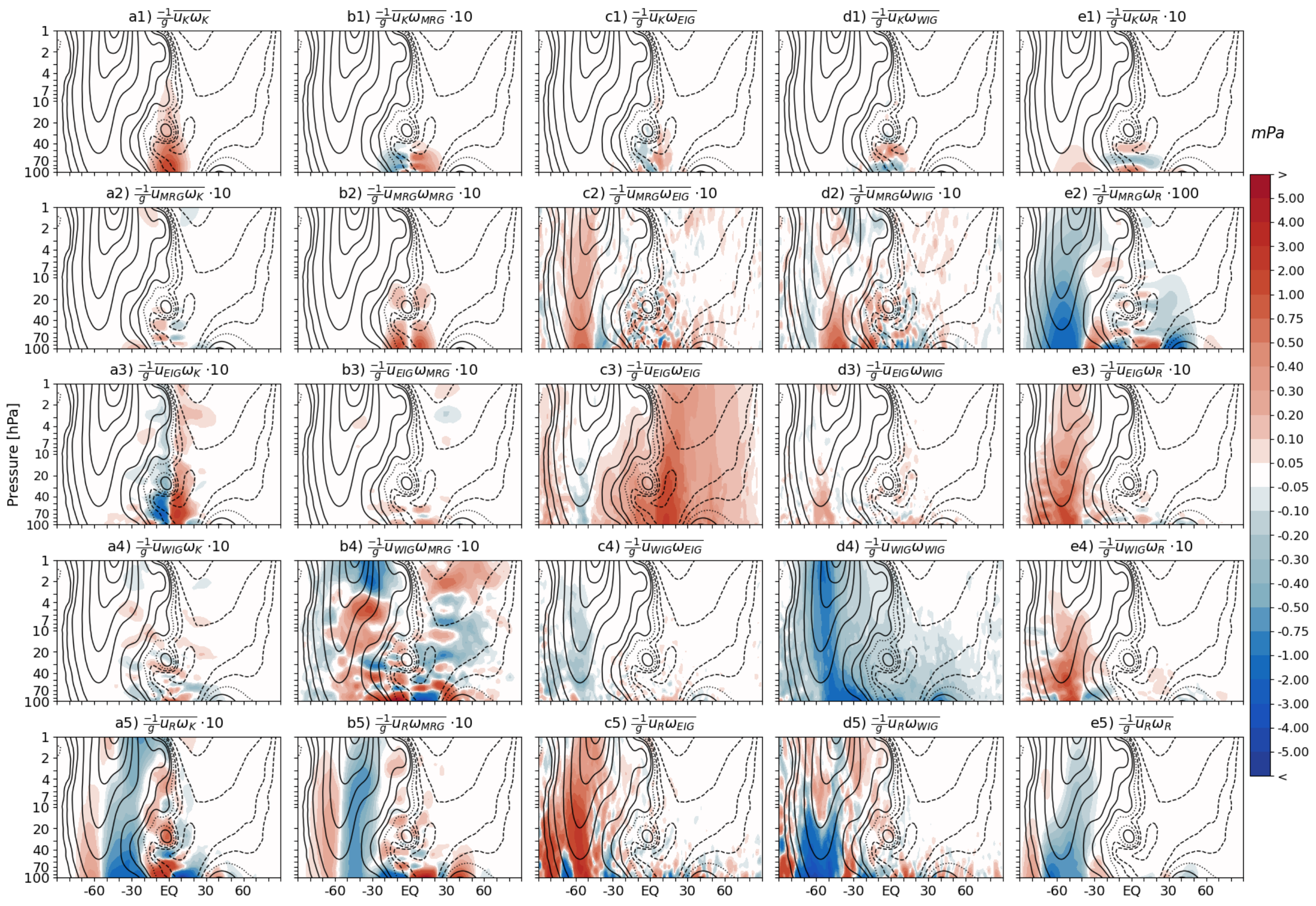}
 \caption{Pressure-latitude distribution of the zonally averaged vertical flux of the zonal momentum (colors) and zonal-mean zonal wind (contour lines) for 00 UTC on 1 July 2020 during the descending westerly shear zone of the QBO.  Fluxes and winds are smoothed by a 5-time-step running mean (from 00 UTC on 30 June until 00 UTC on 2 July). Zonal-wind contours are plotted at intervals of $\pm$10 m/s, with westerlies as solid and easterlies as dashed lines. The zero wind contour is shown by a dotted line. Different panels correspond to different wave types, as defined in the titles and Table~1. }
 \label{fig3}
\end{figure}

\begin{figure}[h]
     \centering
\includegraphics[width=0.99\textwidth]{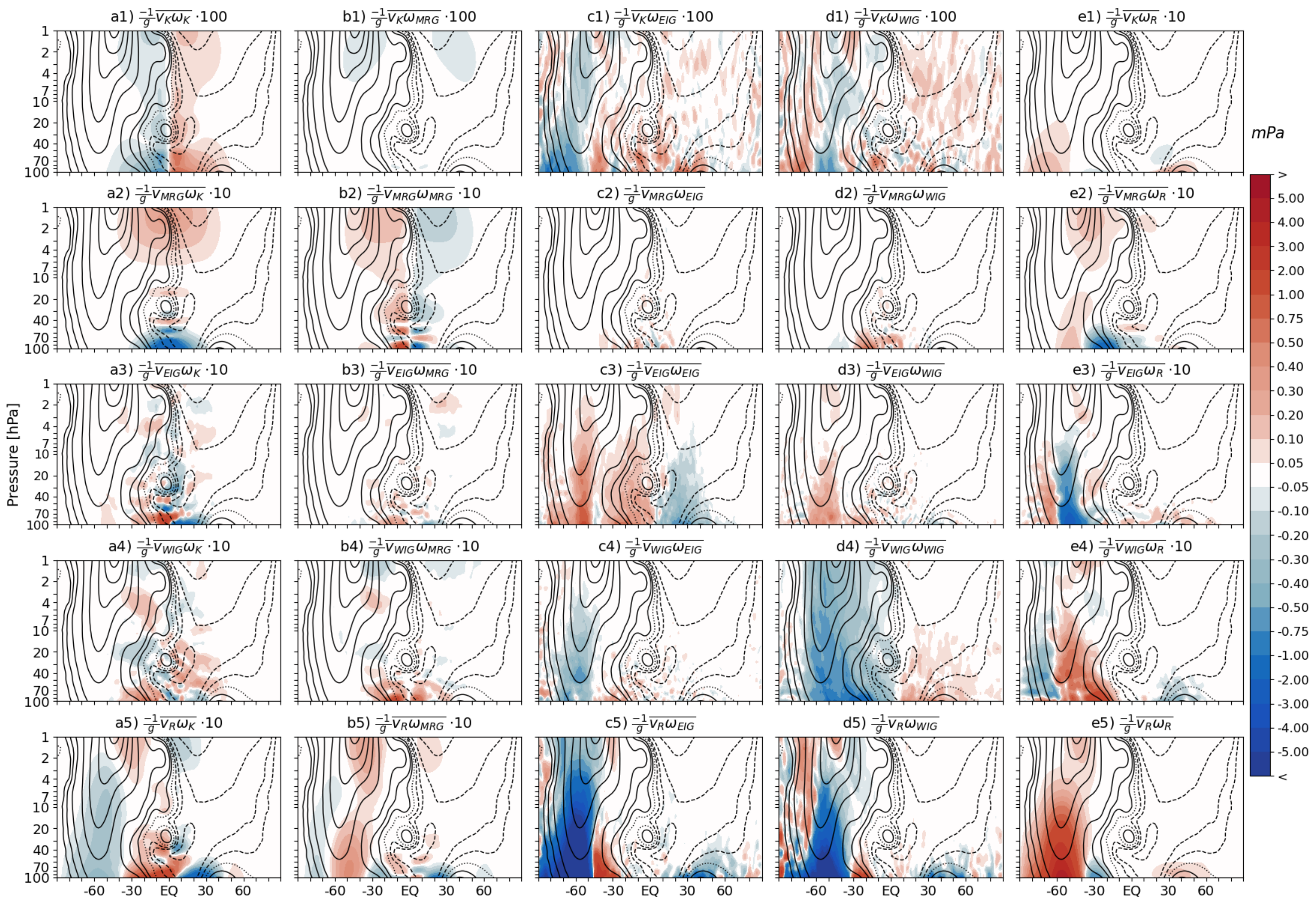}
 \caption{As in Figure~\ref{fig3}, but for the vertical flux of meridional momentum.}
 \label{fig4}
\end{figure}

Instantaneous zonally averaged fluxes on 1 July 2020 are shown in Figures~\ref{fig3}--\ref{fig4} for the VZMF and VMMF, respectively. At that time, the descending westerly QBO (WQBO) jet was centered near 25~hPa, with a strong westerly shear layer between 30 and 40~hPa. In the Southern Hemisphere (SH), the polar night jet was centered near 50$^\circ$S, with westerly winds exceeding 50~m~s$^{-1}$ above 20~hPa. In contrast, the Northern Hemisphere (NH) was characterized by much weaker easterly winds.
Such background flow favors positive (negative) momentum fluxes for waves propagating upward and eastward (westward), respectively, relative to the mean flow \cite<e.g.,>[]{fritts2003gravity}. Positive and negative VZMFs correspond to an upward transport of eastward and westward momentum, respectively. Likewise, positive and negative VMMFs correspond to an upward transport of northward and southward momentum, respectively. 
Note that some flux components in Figures~\ref{fig3}--\ref{fig4} have been multiplied by factors of 10 or 100 so that all components can be displayed using the same colorbar.


\subsection{Diagonal terms of the VMF matrix}

\subsubsection{Kelvin waves} 
The contribution of Kelvin waves (Figure~\ref{fig3},a1) to the VZMF peaks below the WQBO jet with amplitudes between 1-2~mPa within a narrow band around the equator. Outside 10\Deg N-10\Deg S, the Kelvin VZMF falls from about 0.5~mPa at 10\Deg~to less than 0.05~mPa near 18\Deg. The decreasing amplitude of Kelvin VZMF with altitude indicates wave dissipation, momentum deposition, and eastward forcing below the zero wind line, contributing to the downward propagation of the WQBO jet. The VMMF of the Kelvin wave (Figure~\ref{fig4},a1) is negligible as expected since Kelvin waves have very small meridional wind components and vanishing $v'$ at the equator.  

\subsubsection{MRG waves} 
The contribution of MRG waves (Figure~\ref{fig3},b2) to the VZMF has a peak between 5\Deg \, and 10\Deg \, from from the equator, below the WQBO jet. The flux amplitudes are 0.01-0.02~mPa, an order of magnitude smaller than the Kelvin wave flux. This is expected given the smaller magnitudes of the MRG wave zonal winds.  
The positive VZMF decreases with altitude, indicating that MRG waves dissipate, contributing to the WQBO forcing on the flanks. A smaller secondary pair of flux peaks just above the WQBO core might be associated with MRG wave growth due to instability \cite{Mahoetal2025}. 
Both Kelvin and MRG wave VMMFs appear to have a secondary maximum near 1~hPa which may be associated with numerical damping in the IFS model \cite{gupta2024estimates}. 

The spatial structure of the MRG VMMF is more complex (Figure~\ref{fig4},b2); fluxes are asymmetric with several maxima below the WQBO core. The off-\eql flux maxima reflect the fact that the horizontal divergence and thus vertical velocity of MRG waves is zero at the equator \cite{Neduhaletal2024}, despite $v'$ being maximal there.
The small amplitude of the MRG VZMF is in agreement with the results of \citeA{pahlavan2021revisiting} and \citeA{KimChun2015} who argued that most of the MRG wave forcing comes from the meridional fluxes of heat and zonal momentum and not from vertical momentum fluxes.

\subsubsection{ EIG waves} 
The maximum EIG VZMF is found in the lower stratosphere in the summer hemisphere (Fig. \ref{fig3},c3). Its peak amplitude between 2 and 3~mPa, occurs at 10\Deg-15\Deg \, away from the equator. 
This picture is easily understood by considering that EIG waves are defined relative to the state at rest. Since easterly zonal winds dominate the NH stratosphere in summer, EIG waves can propagate vertically without encountering critical levels.  
The opposite is observed in the SH midlatitudes, with little or negative EIG VZMF below the winter polar jet at 50\Deg S-65\Deg S. Within a westerly jet, EIG waves can have westward phase speeds relative to the zonal mean, resulting in negative momentum flux. 

In the tropics, large vertical gradients of the EIG VZMF appear below the WQBO jet in the region of the westerly wind shear. This is the layer where the EIG waves propagating vertically through the region of easterlies reach a critical level and deposit their momentum into the mean flow, forcing the downward propagation of the  WQBO jet \cite{LindzenHolton68}. The VMMF (Figure~\ref{fig4},c3) is negative in the lower stratosphere of the NH, with a peak between 20\Deg N and 30\Deg N, whereas it is positive at winter latitudes in the SH, with maxima between 50\Deg S and 60\Deg S and around 30\Deg S. 

\subsubsection{WIG waves} 
WIG waves exhibit negative VZMF in the stratosphere with larger amplitudes in the winter hemisphere (Figure~\ref{fig3},d4) and a maximum of -1 to -2~mPa in the SH lower stratosphere between 20\Deg-50\Deg. 
Within this region, a narrow belt of nearly vertically invariant negative fluxes of about -0.5-0.75~mPa is located at 45\Deg S-55\Deg S. This is most likely due to orographic gravity waves generated by the Andes hotspot, as described in the ECMWF reanalyses and km-scale simulations by \citeA{gupta2024estimates}. 
In the NH stratosphere, easterlies prevent the vertical propagation of WIG waves. As a result, the VZMF in the NH has a peak peak in the lower stratosphere near 40\Deg N with amplitudes of -0.5 to -0.75~mPa. The VMMF (Figure~\ref{fig4},d4) in the SH is similar, but with a weaker maximum associated with the Andes hotspot. 

In the tropics, the mostly positive VZMF of EIG and negative VZMF of WIG waves partly cancel. WIG waves are responsible for the broad peak of westward momentum fluxes in the winter hemisphere midlatitudes, while EIG waves are responsible for the narrow peak of eastward momentum fluxes in the summer hemisphere subtropics. Therefore, the meridional structure of the combined IG fluxes more closely resembles that reported by studies such as \citeA{wei2022global}, which was derived for gravity waves resolved by the 9-km ECMWF model (figure not shown).

\subsubsection{ Rossby waves} 
The Rossby wave fluxes are shown in panels e5 of Figure~\ref{fig3} and Figure~\ref{fig4}. The VZMF is negative in the winter hemisphere lower stratosphere between 60\Deg S-70\Deg S with an amplitude up to -1~mPa, and is associated with large-scale vertically propagating Rossby waves from the troposphere. In contrast, the VMMF is positive as expected. Because Rossby waves cannot propagate vertically in background easterlies, their VMF is negligible in the summer hemisphere and the tropical stratosphere.

\subsection{Mixed terms of the VMF matrix}

The mixed terms of the $5\times5$ VMF matrix are both puzzling and revealing, as they indicate physical coupling between Rossby and IG modes. 
In the tropics, the mixed terms are generally small, especially the ones involving $\omega_{M}$ and $\omega_{R}$, as seen in the off-diagonal panels in Figures~\ref{fig3}-\ref{fig4}. This is  expected, given the small vertical velocities of MRG and Rossby waves \cite{Zagaretal2023}. Overall, the mixed terms show less coherent spatial structures, which reflects their greater temporal intermittency compared to the diagonal terms. 

The large off-diagonal terms are confined to the high and midlatitudes of the winter hemisphere, involve $\omega_{G}$ and $u'_R$, $v'_R$ and have amplitudes comparable to, or even greater than, those of the diagonal terms. 
They can be understood in the context of the polar night jet, where strong vertical and horizontal shear, together with large potential-vorticity gradients, can modify the propagation of both Rossby and IG waves. 
At large scales, the mixed terms may be related to baroclinic Rossby-wave dynamics and vertical propagation. Their presence may be at least partly associated with the linear decomposition, in which the IG modes represent both gravity waves and low-frequency components that result from the balance between the linear and nonlinear terms of the prognostic equations \cite<e.g.,>[]{Errico1984}. As a result, a large fraction of ageostropphic motions can be projected onto the IG modes, because the Rossby-wave divergence arises only from the beta term \cite<see discussion in>[]{Zagaretal2023}. This may lead to substantial covariance between Rossby-wave horizontal velocities and IG vertical velocities in the winter middle atmosphere, especially in the lower stratosphere. Other processes that may contribute to the observed covariances include local spontaneous generation of IG waves by balanced-jet and frontal dynamics, as well as wave emission from regions of imbalance and linear mode conversion as disturbances propagate through a strongly varying background flow \cite<e.g.,>[]{PlougonvenZhang2014,Dornbarcketal2018,PS2020}. 
In such processes, Rossby and IG waves need not evolve independently, and correlated modal amplitudes can produce finite cross terms, such as $\overline{u'_R \omega'_{G}}$, even though there appears to be a substantial degree of cancellation between $\overline{u'_R \, \omega'_{E}}$ and $\overline{u'_R \, \omega'_{W}}$  (Figure~\ref{fig3},c5-d5). This behavior is further investigated using cross-spectra in the next section. 

\section{Cross spectra and power spectra in the lower stratosphere} 

\subsection{Coherence and phase spectra  }

\begin{figure}[h]
\centering
\includegraphics[width=0.99\textwidth]{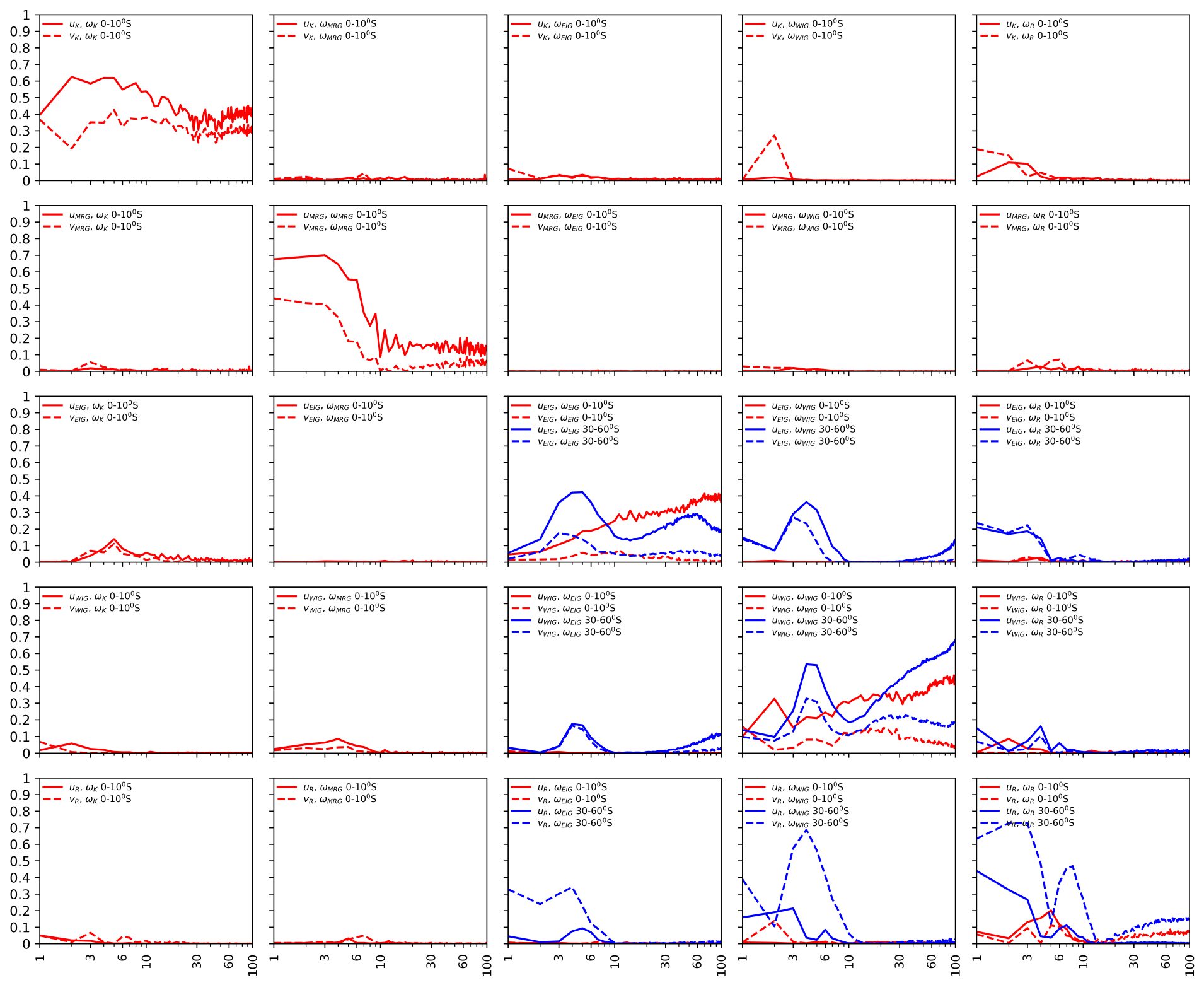} 
 \caption{Zonal wavenumber spectra of the squared spectral coherence between $u'$ and $\omega'$ (full lines) and between $v'$ and $\omega'$ (dashed lines) for five wave types at 80~hPa. Tropical coherences are shown in red, and winter hemisphere midlatitude coherences are shown in blue. Spectra are averaged for June, July, and August (JJA) of 2021.}
 \label{fig5} 
\end{figure}

Figure~\ref{fig5} shows the VZMF (full lines) and VMMF (dashed lines) coherence spectra for the summer months (JJA) of 2021 averaged in latitudinal bands 0\Deg-10\Deg S and 30\Deg S-60\Deg S at 80~hPa. We show only the winter hemisphere, where fluxes due to mixed terms are large.  
The organization of the figure follows Table \ref{T1} with rows associated with the horizontal velocity perturbations and columns associated with $\omega'$ for the five wave types.
The two left columns and the two top rows involving the Kelvin and MRG coherences show only the 0\Deg-10\Deg S band (red curves), whereas other panels also show the 30\Deg S-60\Deg S band spectra (blue). 

\begin{figure}[h]
\centering
\includegraphics[width=0.99\textwidth]{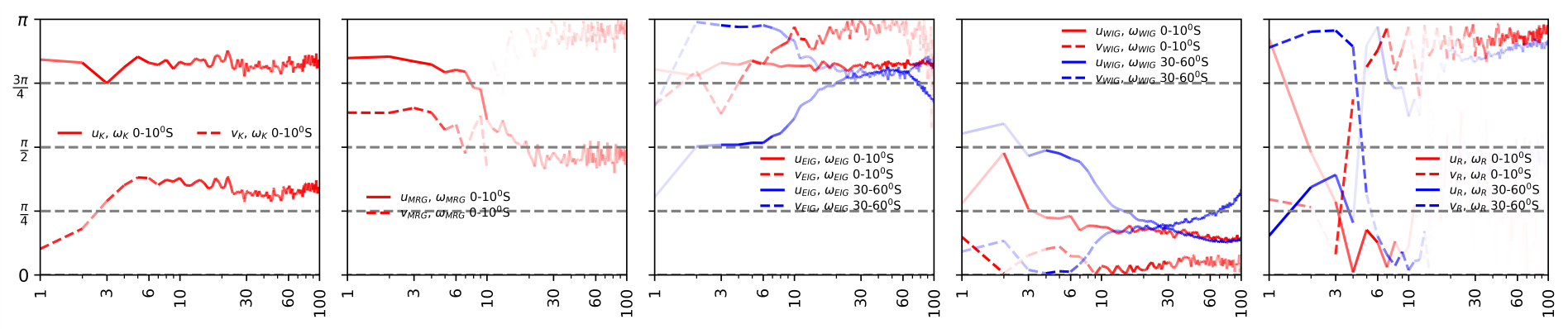} 
\caption{Phase spectra for the $\hat{u}_k$ and $\hat{\omega}_k$ (full lines) and $\hat{v}_k$ and $\hat{\omega}_k$ (dashed lines) in 0-10\Deg S \, band (red curves) and 30-60\Deg \, band (blue curves) for diagonal terms in Figure~\ref{fig5}. The line opacity is proportional to the corresponding fig5.  The five panels correspond to (from left to right) Kelvin, MRG, EIG, WIG and Rossby modes.}
\label{fig6}
\end{figure}

A substantial coherence can be seen in Figure~\ref{fig5} for all diagonal terms, as expected from the relationship between the horizontal and vertical velocities associated with a particular wave mode and waves' common dynamical origin. Phase spectra associated with diagonal terms are shown in Figure~\ref{fig6}. 

There are significant differences in coherence across scales and between the tropics and midlatitudes. 
Kelvin wave fluxes show persistent coherence over all zonal wavenumbers, while MRG-wave fluxes have large coherence at $k<10$ and negligible coherence at  smaller scales. The relative phase shift between $u'_K$ and $\omega'_{K}$ is between  $3\pi/4$ and $\pi$ for all $k$, which means that Kelvin waves at all wavenumbers have a consistent positive contribution to the VZMF in JJA 2021. A similar phase shift is seen for the $u'_M$ and $\omega'_{M}$ at large scales with large coherence.  
Mixed terms involving Kelvin and MRG waves are generally small, suggesting that these terms contribute little to the tropical VMF on monthly and longer time scales. 

Tropical EIG and WIG zonal-wind components show increasing coherence with zonal wavenumber, possibly reflecting the increasing wave-like character of IG modes at smaller scales. The tropical large-scale divergent circulation, such as the Hadley cell and monsoonal flow, projects onto various IG modes,  but it is not necessarily coupled with the large-scale vertical motion, which is generally weak in these systems. The smaller coherence between the IG $v'$ and $\omega'$ may reflect the prevalent zonal movement of the IG modes within the \eql waveguide. Off-diagonal terms have almost no coherence in the tropics, suggesting small contribution to tropical VMFs.

In contrast, mixed terms with EIG, WIG, and Rossby wave velocities in the midlatitude winter stratosphere show some coherence at planetary and synoptic scales, implying a significant contribution to the fluxes at those wavenumbers, in agreement with Figures~3-4. 
The  $u'_E\,\omega'_{E}$ and $u'_W\,\omega'_{W}$  coherence at 80~hPa is characterized by two peaks, one at synoptic scales ($k=4-7$) and another at subsynoptic scales, with a minimum in coherence around $k=10$, suggesting that the two maxima arise from different processes. The former is associated with synoptic-scale ageostrophic circulation projecting on IG modes and can also be seen in $v'_R\,\omega'_{R}$ and off-diagonal terms in midlatitudes. It corresponds to divergent circulation within quasi-geostrophic theory.  
This is also evident from the phase spectra in Figure~\ref{fig6}, where the phase shift between $u'_{E}$ and $\omega'_{E}$ at $2<k<10$ is around $\pi/2$, while the phase shift between and between $v'_{E}$ and $\omega'_{E}$ at $2<k<10$ is around $\pi$. A similar $\pi/2$ difference in phase shift between the two wind components can be seen for WIG modes at large scales. 
Increasing coherence at subsynoptic scales, especially for $u'_W\,\omega'_{W}$ is likely associated with  IG wave dynamics. 

\subsection{Tropical VMF cospectra at 80~hPa}

\begin{figure}[ht]
\centering
\includegraphics[width=0.99\textwidth]{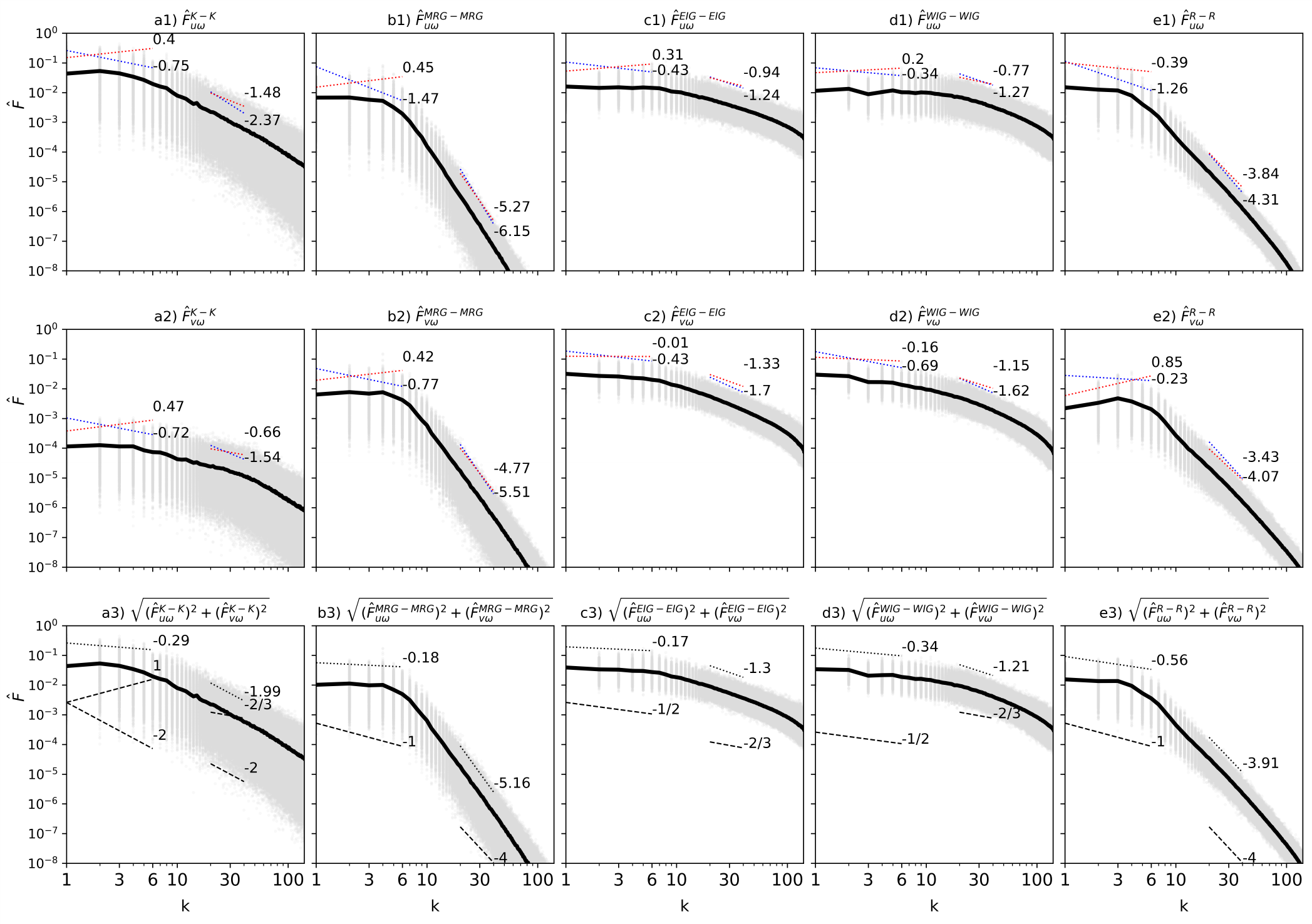} 
\caption{Zonal wavenumber cospectra of the VZMF (a1-e1), VMMF (a2-e2) and absolute VMF (a3-e3) for diagonal terms of the flux matrix. Five columns (from left to right) correspond to the Kelvin, MRG, EIG, WIG, and Rossby wave flux cospectra. Grey circles are cospectra at individual time steps (00~UTC and 12~UTC during 1 July 2019 to 30 June 2021) and their two-year averages are shown as thick black lines. All cospectra are averaged over the 10\Deg N-10\Deg S band at 80~hPa. Thin red (blue) dotted lines are associated with the largest (smallest) slopes of the monthly averaged spectra at large scales ($k=1-6$) and at subsynoptic scales ($k=20-40$). Black dashed and dotted lines in the bottom row are theoretical and empirical slopes, respectively, for the power laws of the absolute VMF cospectra.}
\label{fig7} 
\end{figure}

Now we investigate the extent to which the VMF in the ECMWF analyses is consistent with the scaling relationships summarized in Table~\ref{T3}. This is examined using two-year mean cospectra of the VZMF, VMMF, and absolute VMF for the diagonal flux terms, shown in Figure~\ref{fig7}. 

The largest spread in the cospectra over the 2-year period can be seen in Figure~\ref{fig7} for Kelvin wave, indicating their largest temporal variability.  
At planetary scales, the Kelvin-wave VMF cospectra can even exhibit positive slopes, but the two-year mean slope of approximately ~-1/3 (panel a3) falls within the range expected from Table~\ref{T3}. At synoptic scales, the slope is very close to the steeper limiting value of $-2$ expected for Kelvin waves. 

For MRG waves, the HKE spectrum has a relatively flat slope at large scales but follows a -3 power law at  subsynoptic scales, as shown by \citeA{Neduhaletal2024} (see also Figure~1a). This implies VMF cospectral slopes between $-2$ and  $-1$ at planetary scales and a slope of $-4$ at subsynoptic scales (Table~2). Our VMF data show nearly flat cospectra at planetary scales, indicating stronger fluxes at these scales than expected from the theoretical scaling, whereas the approximately $-5$ power law at subsynoptic scales suggests a more rapid decrease in VMF than expected over this range of wavenumbers (panels b1-b3). 

For Rossby waves, the subsynoptic-scale VMF cospectrum approximately follows the expected $-4$ power law, similar to the HKE. The tropical planetary scale VMF cospectrum is expected to follow a $-1$ power law, just like the HKE spectrum. The Rossby mode cospectra in panels e1-e3 of Figure~\ref{fig7} are somewhat shallower, just like the HKE spectrum in the lower stratosphere in Figure~1a.   

Similarly, the relatively flat planetary-scale HKE spectra for the IG modes (Figure~1a), associated with the large-scale unbalanced circulation, produce relatively flat VMF cospectra (panels c1-c3, d1-d3) with slopes lying in the expected range between $-1/2$ and $1/2$ (Table~2). At the largest subsynoptic scales, the theoretically expected slopes around $-1$ are found in the EIG modes, whereas the absolute VMF cospectrum is steeper. 
For $k>100$, the VMF cospectra decrease further in amplitude, reflecting the lack of IG variability in the data. 

Figure~\ref{fig7} as a whole illustrates that IG modes dominate the subsynoptic-scale VMF, whereas  the large-scale VMF in the tropical lower stratosphere is contributed by various modes, in agreement with previous studies \cite<e.g.,>[]{KimChun2015,pahlavan2021revisiting}, but without imposing predefined scale and frequency cutoffs and allowing different wave modes to coexist across scales.

\section{VMFs in the two phases of the QBO} 

\begin{figure}
\centering
\includegraphics[height=0.838\textheight]{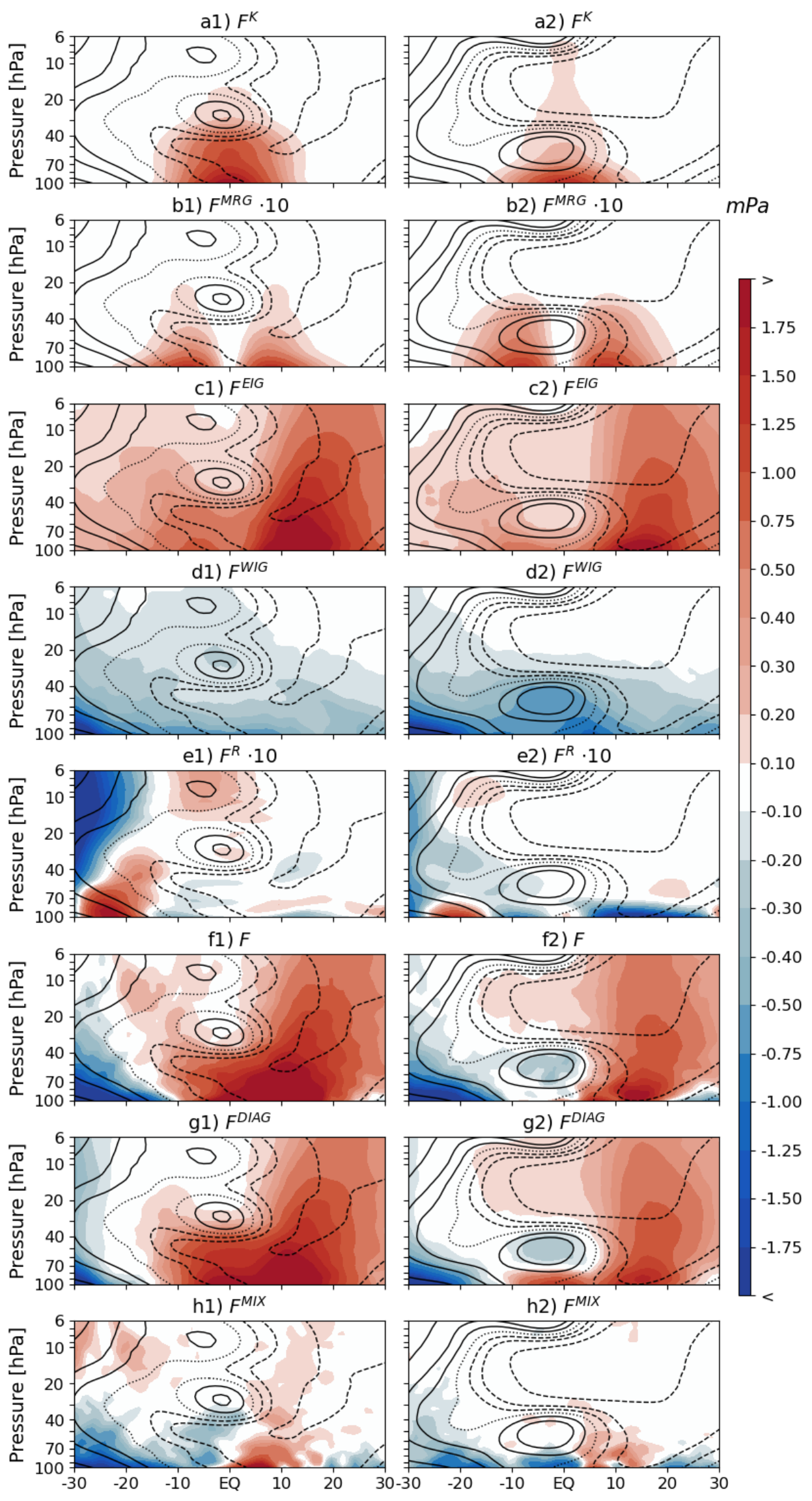} 
\caption{Latitude--distribution of VMFs (colors) and zonal-mean zonal winds (lines) for the descending (panels a1 through h1 in the left column) westerly and (panels a2 through h2 in the right column) easterly shear zones of the QBO. The averaging period is 26 May--24 August 2020 for the descending westerly and 22 March--20 June 2021 for the descending easterly shear zone. The isoline spacing for the zonal wind is the same as in Figure~\ref{fig9}. The MRG and Rossby wave VMFs are multiplied by 10.}
\label{fig8} 
\end{figure}

This section compares the zonally averaged VZMFs between the westerly and easterly phases of the QBO. 
Periods of descending westerly and easterly shear zones (WSZ and ESZ, respectively) are defined as periods during which the 90-day average zero-wind line of the zonal-mean zonal wind at the equator is closest to 35~hPa, with westerly winds above and easterly winds below the zero-wind line for WSZ, and vice versa for ESZ. The two periods are boreal summer 2020 (26 May--24 August) and boreal spring 2021 (22 March--20 June) for WSZ and ESZ, respectively. 
The latitude--altitude distributions of fluxes and zonal-mean zonal winds for the  WSZ and ESZ periods are shown in the left and right columns of Figure~\ref{fig8}, respectively. 

\begin{figure}[h!]
\centering
\includegraphics[width=0.99\textwidth]{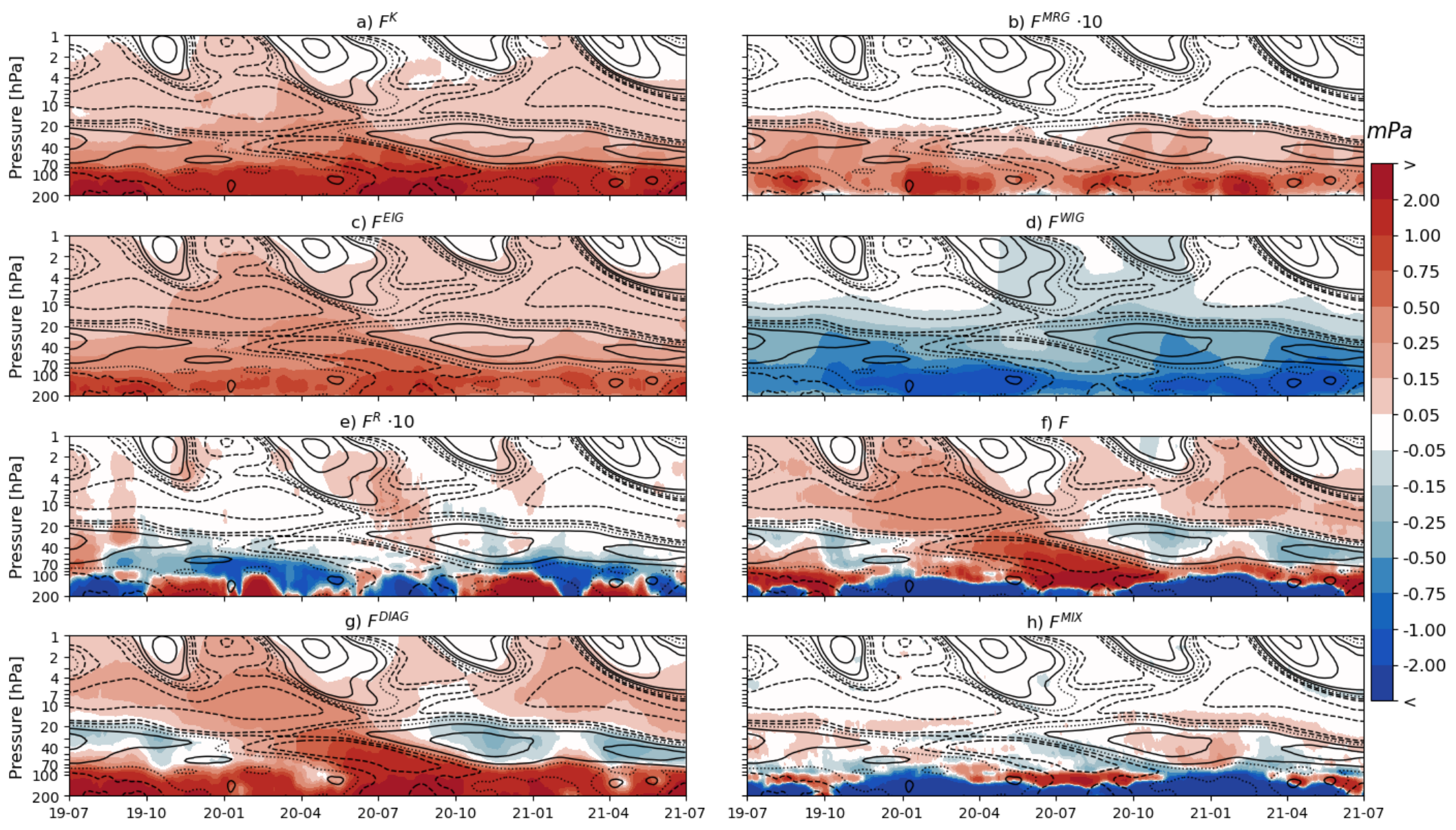} 
\caption{Time--altitude cross section of the VMFs associated with the diagonal terms of a) Kelvin, b) MRG, c) EIG, d) WIG and e) Rossby modes, f) total VMF, and VMF due to g) all diagonal terms and h) all mixed terms. The fluxes are averaged over 5$^{\circ}$S--5$^{\circ}$N and additionally smoothed using a 30-day moving mean. Isolines show the zonal-mean zonal wind, with easterlies shown by dashed lines (-50, -30, -10, -5 m~s$^{-1}$), westerlies by solid lines (5, 10, 20, 30, 40, 50 m~s$^{-1}$), and the zero-wind line shown as dotted. Fluxes of MRG and Rossby waves are multiplied by 10. Labels on the x-axis show the year and month as yy-mm between July 2019 and June 2021.}
\label{fig9}
\end{figure}

To support the interpretation, we show in Figure~\ref{fig9} the temporal evolution of the  different terms of the flux matrix between July 2019 and June 2021 from 200~hPa to 1~hPa, together with the zonal-mean zonal wind. This figure shows that the total positive flux within the tropical tropopause layer is dominated by the diagonal terms (Figure~\ref{fig9}g), while the negative flux below 100~hPa is dominated by the mixed terms (Figure~\ref{fig9}h). Strong positive fluxes can be seen in regions of easterlies during boreal spring and summer of 2020, whereas negative fluxes occur at other times in the lower stratosphere (Figure~\ref{fig9}f,g). 

\begin{figure}
\centering
\includegraphics[width=1.0\textwidth]{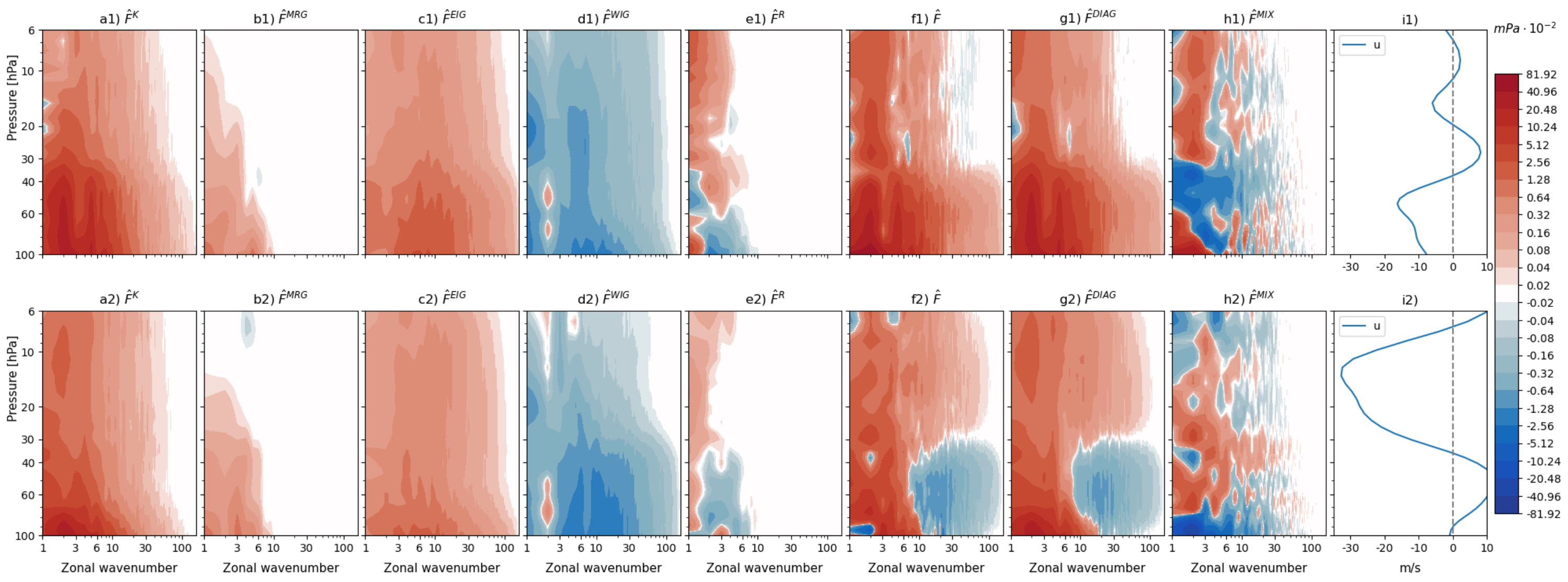} 
\caption{Altitude--zonal wavenumber distribution of VMFs for the descending westerly (panels a1 through h1 in the top row) and easterly (panels a2 through h2 in the bottom row) shear zones averaged between 5$^{\circ}$S--5$^{\circ}$N. Panels i1) and i2) show the vertical profiles of the zonal-mean zonal wind, averaged over the same period and latitude band.}
\label{fig10}
\end{figure}

The zonal wavenumber VMF spectra for different elements of the flux matrix are shown in Figure~\ref{fig10} for the two periods shown in Figure~\ref{fig8}. The spectra quantify the contributions of various waves at different scales, and their vertical variations are discussed in relation to the vertical profiles of the zonal-mean zonal wind. As before, we discuss the results in terms of the diagonal components of the modal flux matrix, the mixed terms, and the total VMF. 

\subsection{Kelvin waves}

The Kelvin wave VMF is largest during the June-November period, when prevailing easterly winds in the troposphere provide a favorable environment for the vertical propagation of Kelvin waves into the stratosphere (Figure~\ref{fig9}a). When the QBO easterlies reach the tropopause, such as during June-November of 2020, Kelvin waves can propagate higher up into the stratosphere. Their VMF amounts to $\approx$2~mPa  at the equator during both phases. In the \bkg easterlies, fluxes slowly decline with altitude until reaching the altitude of the WSZ (Figure~\ref{fig9}a and Figure~\ref{fig8}a1). Just below the zero wind line in the WSZ, at around 40~hPa pressure level, Kelvin wave fluxes are $\approx$1~mPa.  Within the WSZ, Kelvin waves dissipate their momentum and are absorbed at their critical level. The deposited eastward momentum provides eastward forcing of the mean flow, effectively causing the downward propagation of the WSZ. This is evident from the rapid decline of Kelvin wave fluxes in the WSZ near the zero wind line in Figure~\ref{fig8}a1. 

The scale distribution of Kelvin wave VMFs for the WSZ and ESZ periods shows that the largest flux contributions come from the large scales, $k=1-6$ (Figure~\ref{fig10}, a1-a2). There is a rapid reduction of VMF with $k$, so that the contribution at $k=10$ is 2--4 times smaller whereas the contribution at $k=30$ is 16--32 times smaller than those at large scales. Panel a1 of Figure~\ref{fig10} also shows that Kelvin wave fluxes stay nearly constant within the 50-100~hPa layer of easterlies while the fluxes rapidly decline with height at all scales in the WSZ within the 50-30~hPa layer. In the ESZ, Kelvin wave fluxes between 50~hPa  and 10~hPa are nearly constant across scales, implying vertical propagation without significant dissipation (Figure~\ref{fig10}, a2). 

The scale dependency of Kelvin wave VMFs suggests that significantly larger fluxes obtained in the present study, compared to those in earlier studies \cite{ern2009,tindall2006equatorial,ern2009quantification}, may be related to the applied Kelvin wave cutoff scale, in addition to other factors. For example, \citeA{ern2009quantification} used $k=6$ as the cutoff scale, neglecting the Kelvin wave fluxes on synoptic and subsynoptic scales. We shall quantify the role of the cutoff scale in the next section. 

\subsection{MRG waves}

MRG wave VMF evolution is shown in panel b) of Figure~\ref{fig9}, and the two QBO phases are compared in panels b1) and b2) of Figure~\ref{fig8} and Figure~\ref{fig10}. Figure~\ref{fig8} shows that the MRG wave fluxes in the lowermost stratosphere have a maximum of $\approx$0.1~mPa off the equator at around 8\Deg N and 8\Deg S. This is because the VMF is proportional to the zonal wind, which is zero at the equator, and whose maximum depends on the MRG wave scale \cite{Mahoetal2024}.
This latitude range is not included in the other two figures, where averaging over 5\Deg N--5\Deg S was used for comparison with previous studies.

The largest vertical decline in the MRG VMF occurs at the zero-wind line in the ESZ, where, by dissipating their eastward momentum, MRG waves cause eastward acceleration on the flanks of the westward QBO (Figure~\ref{fig8},b2) and below the WSZ on the flanks of the westward QBO (Figure~\ref{fig8}, b1). Qualitatively similar results were obtained by \citeA{pahlavan2021revisitingPART2}. It is important to note that the MRG VZMF accounts for only a minor part of the total MRG wave forcing in the QBO region; the bulk of the MRG forcing is due to the equatorward or poleward flux of zonal momentum (primarily in the WSZ) and the equatorward or poleward heat flux (primarily in the ESZ) \cite{pahlavan2021revisitingPART2}.

Figure~\ref{fig10} also shows that nearly all MRG fluxes are contained within the first 10 zonal wavenumbers. The peaks in the fluxes at $k=6$ and $k=3$ in the WSZ and ESZ, respectively, may be associated with the mechanism discussed by \citeA{Mahoetal2025}.

\subsection{EIG waves}

The EIG fluxes have maximum of $\approx$2~mPa  at around 12$^{\circ}$N in the lowermost stratosphere. The maximum shifts somewhat poleward with altitude so that the largest EIG fluxes at 10~hPa can be found around 20$^{\circ}$N (Figure~\ref{fig8}, c1-c2). 
This peak is associated with the tropical convection \cite<e.g.,>[]{geller2015tropical,polichtchouk2023resolved} and is therefore seasonal, with a smaller peak during December-February around 12$^{\circ}$S (not shown). 

Within the  5\Deg N--5\Deg S belt, the EIG fluxes are between 0.50~mPa and 0.75~mPa below the zero wind line in the WSZ (Figure~\ref{fig8},c1). Near the zero wind line they dissipate and cause the eastward acceleration and downward propagation of the westerly QBO jet. 
Above the zero wind line, in the westerly QBO jet, EIG fluxes are between 0.10~mPa and 0.20~mPa, which implies that up to 0.65~mPa of fluxes have dissipated in the WSZ. 
In the ESZ (Figure~\ref{fig8}c2), EIG waves that propagate into the stratosphere continue upward until they reach their critical level near 7~hPa, where they dissipate near the zero-wind line and cause the eastward acceleration and downward propagation of the next westerly QBO jet. 

The largest EIG fluxes are at scales with wavenumbers between $k=3$ and $k=30$ (Figure~\ref{fig10}c1 and c2), but all scales contribute to the downward forcing of the westerly QBO, as indicated by the vertical gradient of fluxes at all scales in Figure~\ref{fig10},c1 between 50 and 30~hPa. In the ESZ (Figure~\ref{fig10},c2), EIG fluxes are nearly constant with altitude, which indicates that EIG waves do not dissipate and therefore do not contribute to the downward forcing of the easterly QBO.

\subsection{WIG waves}

The WIG fluxes show the opposite behavior to the EIG fluxes. A peak of -1.25~mPa can be found around 5$^{\circ}$N in the lowermost stratosphere (Figure~\ref{fig8},d2) when the eastward QBO winds reach the tropopause. Averaged around the equator, below the zero wind line in the ESZ, WIG fluxes are between -0.50 and -0.75~mPa and between -0.10 and -0.20~mPa above (Figure~\ref{fig8},d2 and Figure~\ref{fig9}d), which implies that up to 0.65~mPa of flux has dissipated in the ESZ. 

The largest WIG fluxes are from waves with zonal wavenumbers between $k=3$ and $k=30$ (Figure~\ref{fig10}d1 and d2), but all scales contribute, similar to the EIG wave fluxes. 
In the WSZ (Figure~\ref{fig10},d1), WIG fluxes are nearly constant with altitude, which indicates that WIG waves do not dissipate and therefore do not contribute to the downward forcing of the westerly QBO. The reason for  positive WIG fluxes at wavenumber 2 is unknown and this feature may also be an artifact of the data assimilation. More substantial dissipation of the WIG waves in the ESZ than of the EIG waves in the WSZ can be attributed to the stronger wind filtering extending to higher frequencies for the WIG waves as a consequence of the higher zonal wind speeds in the easterly QBO jet above the zero wind line in the WSZ case (Figure~\ref{fig10},i1-i2).

\subsection{Rossby waves}

Fluxes of Rossby waves are in the tropical stratosphere are small (Figure~\ref{fig9}e) in both the WSZ (Figure~\ref{fig8},e1) and the ESZ (Figure~\ref{fig8},e2). They exhibit positive and negative (eastward and westward) fluxes that amount to -0.02~mPa below, and 0.02~mPa above the zero wind line in the WSZ (Figure~\ref{fig8}e1), providing westward forcing and slowing down the downward propagation of the westerly QBO jet. In the ESZ, Rossby fluxes are even smaller and contribute little to the downward propagation of the easterly QBO. Nearly all Rossby fluxes are contained within the first 6 zonal wavenumbers (Figure~\ref{fig10},e1 and e2).

\subsection{Diagonal vs. mixed terms} 

Summing the contributions of the five individual flux terms gives the flux associated with the diagonal terms, which can be seen in Figure~\ref{fig9}g, and panels g1 and g2 of Figure~\ref{fig8} and Figure~\ref{fig10}.  The sum of all mixed terms can be seen in Figure~\ref{fig9}h, and panels h1 and h2 of Figure~\ref{fig8} and Figure~\ref{fig10}. The sum of mixed and diagonal fluxes gives the total flux shown in Figure~\ref{fig9}f, and panels f1 and f2 of Figure~\ref{fig8} and Figure~\ref{fig10}. 

These figures show that total flux in the tropical stratosphere is well represented by the diagonal terms. The diagonal terms capture the main features of the VMF and the differences between the WSZ and ESZ associated with critical-level filtering. The mixed terms are substantially smaller, reaching maximum amplitudes of about 0.15~mPa in the ESZ and -0.25~mPa in the WSZ, and are concentrated near the zero-wind line. Their contribution to the total VMF in the QBO region is therefore secondary. Although the mixed terms may contain information about interactions between different wave types, their detailed interpretation is beyond the scope of this study.
 
\section{Sensitivity of VMF estimates to the cutoff scale}

\begin{figure}
\centering
\includegraphics[width=0.99\textwidth]{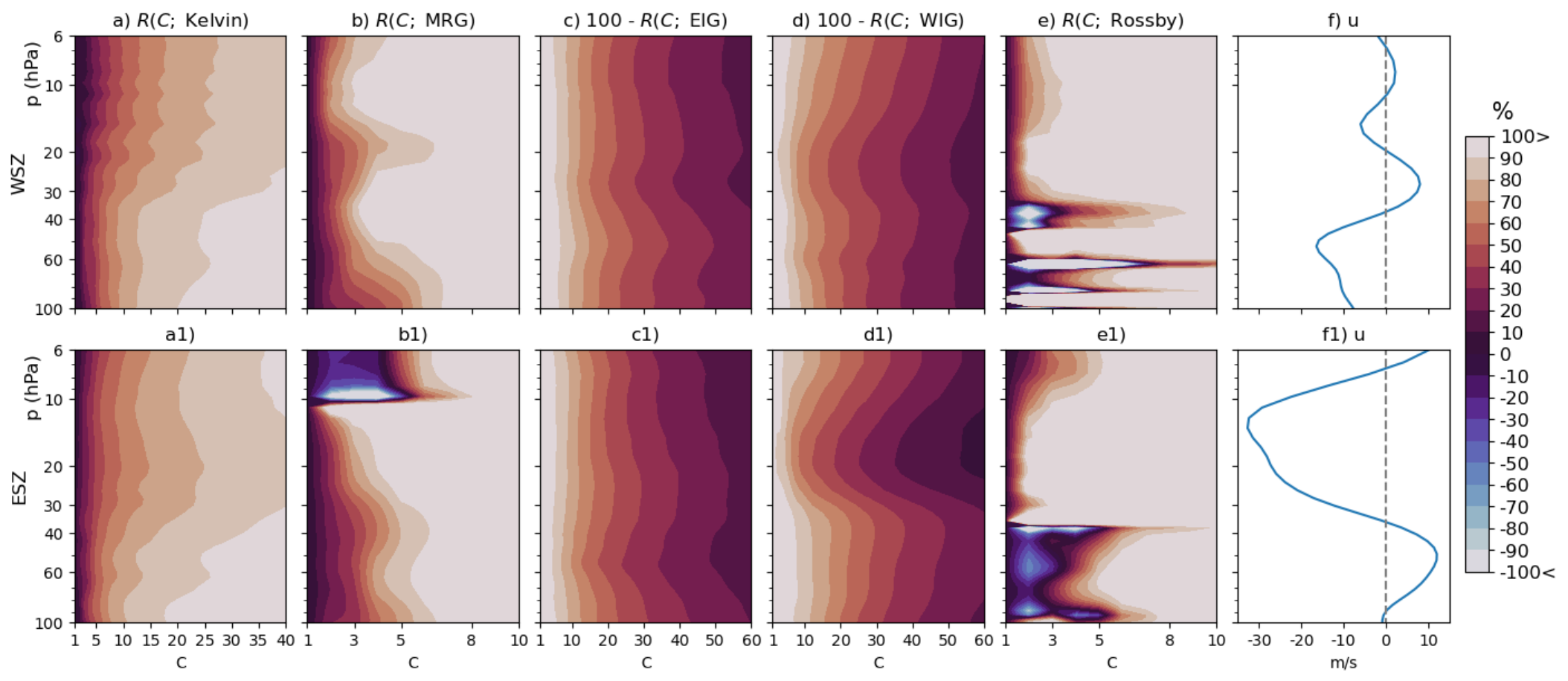} 
\caption{Scale contributions of a)-a1) Kelvin, b)-b1) MRG, c)-c1) EIG, d)-d1) WIG and e)- e1) Rossby wave VMF to the total flux of the same wave type. Relative contributions are computed from cospectra averaged between 5$^{\circ}$N-5$^{\circ}$S during the WSZ period July 2020 (top row) and the ESZ period June 2021 (bottom row). Panels f)-f1) show the zonal-mean zonal wind profiles averaged for the same periods and latitude band.}
\label{fig11} 
\end{figure}

What is the effect of the zonal wavenumber cutoff scale on the estimated VMF? 
To quantify this effect, we compute the cumulative contribution of zonal wavenumbers up to the cutoff scale $C$ and normalize it by the total flux associated with the wave type $j$:
\begin{equation}
\label{Rx}
R(C; \varphi, p, j) = 100\frac{ \sum_{k=1}^C \Re \Big( \hat{F}(k;\varphi,p,j) \Big)}{ \sum_{k=1}^K \Re\Big(\hat{F}(k; \varphi, p, j)\Big)} \, .
\end{equation} 
The quantity $R(C;j)$ measures the percentage contribution of the first $C$ zonal wavenumbers to the total flux. The results for the five wave types are presented in Figure~\ref{fig11}, for the two periods with WSZ and ESZ discussed in the previous section. 

Figure~\ref{fig11} (panels a and a1) shows that the first 10 zonal wavenumbers account for around 70-80\% of the Kelvin VMF in the WSZ period within the 40-100~hPa easterlies (Figure~\ref{fig11}f). As Kelvin waves dissipate and their fluxes decreases around the zero-wind line, the cutoff scale required to capture 80\% of the VMF increases with height. At 10~hPa, the first 25 zonal wavenumbers must be considered. In the ESZ, 80\% of the Kelvin wave VMF is contained within the first 10 wavenumbers below 60 hPa. Above this level, within the strong easterlies in the $15-30$~hPa layer, 80\% of the Kelvin VMF is contained within the first 20 wavenumbers. Thus, using $C=20$ as a zonal wavenumber cutoff for Kelvin wave VMF \cite<e.g.,>[]{pahlavan2021revisitingPART2} leads to capturing up to 90\%  of the flux below the QBO shear zone, and up to 80\% above the WSZ. The exact percentage of  Kelvin wave VMF captured by the wavenumber-frequency method will also depend on the frequency cutoff scale used to identify Kelvin waves. The difference in the estimated Kelvin VMF between the wavenumber-frequency and the new approach will also be affected by the method used to filter $\omega'_K$ and $u'_K$.

Panels b and b1 in Figure~\ref{fig11} show that the first 8 zonal wavenumbers contain nearly all MRG fluxes in both shear zones and at all levels. Negative percentages above 10~hPa and for $C=2-5$ in ESZ appear because at those altitudes and scales, MRG waves have a negative contribution to the total MRG flux, while total MRG flux is positive (see also Figure~\ref{fig10}, b2). The same applies to Rossby VMF which is  almost entirely contained within the first 10 zonal wavenumbers (Figure~\ref{fig11},e and e1) in both shear zones. Below the zero wind lines, Rossby wave fluxes tend to be negative (Figure~\ref{fig10},e1 and e2) while above the zero-wind line the cumulative contribution is positive, with nearly 100\% of the  Rossby VMF captured by $C$ as small as 5.  

For IG modes, we display $100-R(C)$ instead of $R(C)$ to present the cumulative contribution of scales smaller than the cutoff scale ($k>C$). 
Panels c and c1 of Figure~\ref{fig11} show that 50-60\% of EIG VMF in both shear zones originates from zonal wavenumbers greater than $20$, whereas 40-50\% originates from $k<20$.  The first 10 zonal wavenumbers account for  20-30\% of EIG fluxes. The WIG VMFs show a similar distribution in the WSZ, while in the ESZ, above the zero wind line (30-10~hPa in Figure~\ref{fig11},d1) only 30-40\% of the total WIG VMF is accounted for by $C=20$. 

\begin{figure}
\centering
\includegraphics[width=0.8\textwidth]{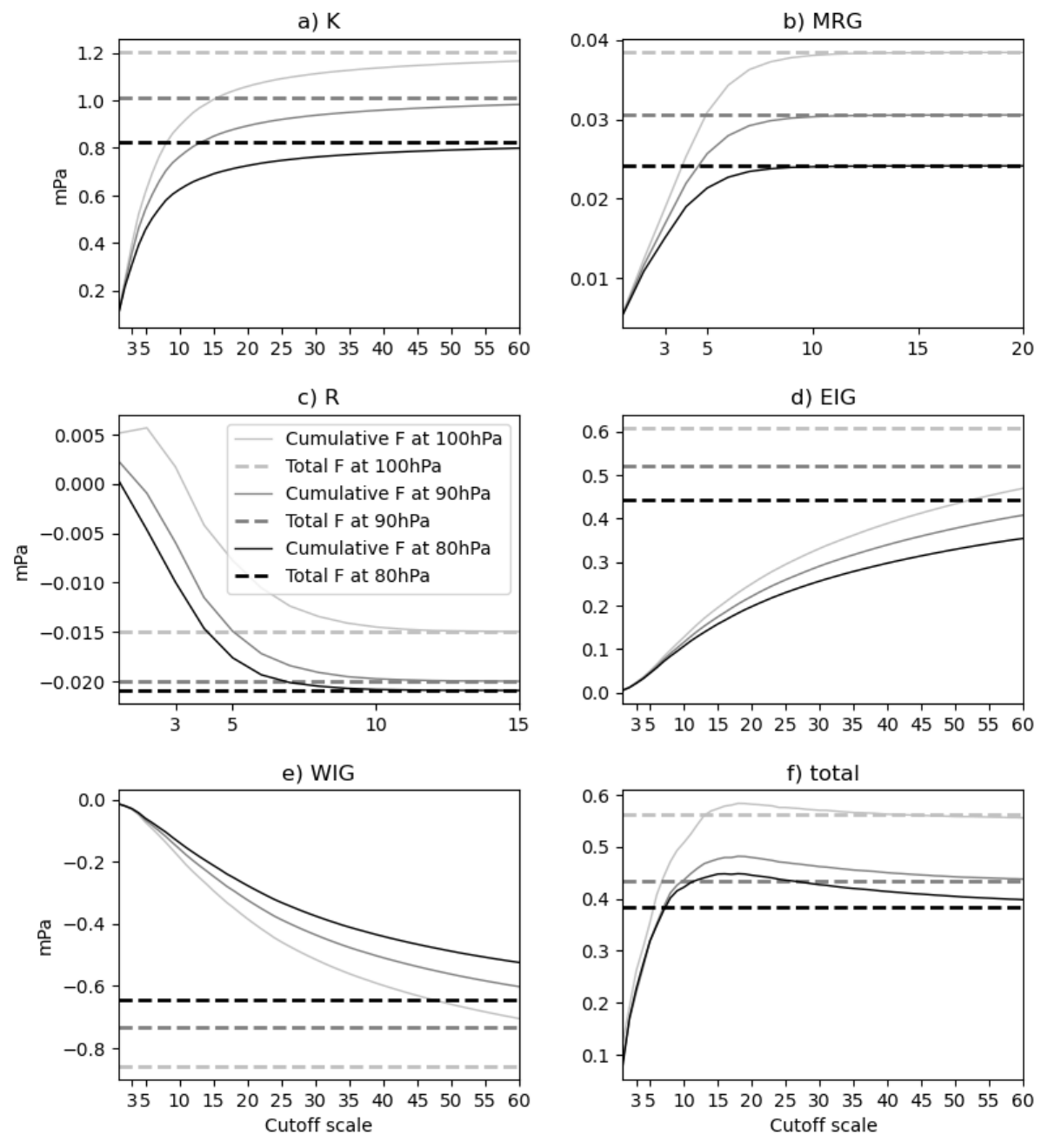} 
\caption{Cumulative VZMFs as a function of the cutoff \wn for various wave types at 100~hPa, 90~hPa and 80~hPa. The total flux in the first 160 wavenumbers is shown by the horizontal dashed lines. a) Kelvin, b) MRG, c) EIG, d) WIG, e) Rossby and f) total VMF. The VZMF is averaged over the 2-year period, July 2019--June 2021 and over latitudes 5\Deg S-5\Deg N.}
\label{fig12}
\end{figure}

Integrated VZMFs up to a given cutoff \wn are shown in Figure~\ref{fig12} for various wave types at 100~hPa, 90~hPa and 80~hPa (three shades of gray),  averaged over the full 2-year period and the 5\Deg S--5\Deg N \, band.  We focus on the lower stratosphere to show the fluxes entering the QBO region as a function of the zonal wavenumber cutoff scale. The corresponding figure showing the fluxes normalized by the total flux is provided in the SI. 
The cumulative and total VMFs at the three levels suggest that the waves that contribute most to the QBO forcing, Kelvin, EIG and WIG modes, have similar cumulative distributions below the QBO region, associated with their vertical propagation and gradual dissipation. This is confirmed by the normalized cumulative fluxes in Figure~S6 of the SI. 
 
At large scales, the magnitude of the Kelvin VMF substantially exceeds the IG VMF (Figure~\ref{fig12}). This contrasts with the results \citeA{sato1997estimates}, who suggested that Kelvin and IG waves have comparable VMF magnitudes, based on twice-daily rawinsonde data at Singapore, but without quantifying their contributions by zonal scale. Together with the numerical simulations of \citeA{Dunkerton1997}, these studies established the role of small-scale gravity waves in driving the QBO. Our study  suggests that, in the current ECMWF analyses, the majority of the VMF at the levels 100--80~hPa occurs at zonal scales larger than 2000~km. Using smaller scales than 100~km would not substantially change this conclusion, given the VMF cospectra and cumulative distributions. 

In contrast to the other modes, Rossby waves on average have similar VMF values at 100~hPa and 90~hPa, but their flux within the 5\Deg S-5\Deg N region is small (Figure~\ref{fig12}c). Here we note that lateral forcing associated with baroclinic Rossby waves propagating from the winter hemisphere would partly project onto synoptic-scale IG modes in the subtropics in our linear decomposition. This process may contribute to the maximum of the IG VMFs in the subtropical lower stratosphere in Figure~8. Lateral forcing was not considered by \citeA{Dunkerton1997}, although that study noted that Rossby waves from the winter hemisphere may be important in the upper and lowermost parts of the QBO and in the subtropics.

\section{Conclusions and Outlook}

We derived expressions for the vertical momentum flux (VMF) associated with Rossby waves, eastward- and westward--propagating inertia--gravity waves (EIG and WIG waves, respectively), Kelvin waves and mixed Rossby–gravity (MRG) waves for a spherical, hydrostatic atmosphere. 
The spherical approach unifies theories on the midlatitude $f$- or $\beta$-plane and on the equatorial $\beta$-plane applied in previous studies. In the new framework, five wave species, defined in terms of the eigensolutions of the primitive equations linearised around the state of rest, coexist across the spatio-temporal scales represented by 3D atmospheric data. In this linear approach, IG modes represent both IG waves and gravity waves unaffected by rotation, as well as slowly varying large-scale divergent circulation other than the Kelvin and MRG waves. The new VMF decomposition removes the need to impose a cutoff wavenumber in the extratropics and for cutoff frequency and wavenumber criteria in the tropics. 

The wavenumber decomposition of the horizontal and vertical velocities into the five wave components thus yields 25 components of the VMF for each horizontal velocity component. This $5\times 5$ VMF matrix has not been discussed previously. The diagonal elements of the flux matrix, in which horizontal momentum is transported vertically within the same wave type, are shown to dominate in the tropical stratosphere. In the extratropical stratosphere, the off-diagonal (or mixed) terms are important. Their presence may partly be associated with the linear decomposition,  but it could also be related to IG modes being locally generated, rather than propagating from the troposphere. A further investigation of the mixed terms in the winter hemisphere stratosphere should be undertaken in future studies. 
 
The analytically derived expressions couple the power laws for the VMF cospectrum, $F_k$, and the mechanical energy spectrum, $I_k$, for every zonal wavenumber $k$; 
$F_k \propto \nu_k I_k $, where $\nu$ is the normal mode frequency. Invoking the frequency relationships for the Rossby and inertia–gravity modes and the $k^{-3}$ and $k^{-5/3}$ power laws for $I_k$ within the inertial range, the VMF cospectra follow $k^{-4}$ and $k^{-2/3}$ power laws for the Rossby and inertia–gravity modes, respectively. In the tropical stratosphere, VMF power-law scalings vary from 
$F_k \propto k^{-1/2}-k^{1/2}$ and $F_k \propto k^{-1}-k^0$ at planetary scales to $F_k \propto k^{-1}$ and $F_k \propto k^{-2}-k^{-1}$ at synoptic and subsynoptic scales of the IG modes and Kelvin waves, respectively. 
The planetary-scale MRG and Rossby wave VMF follow $F_k \propto k^{-2}-k^{-1}$ and $F_k \propto k^{-1}$ power laws, respectively. 
These power laws represent a methodological step forward relative to previously used power laws for the total VMF \cite<e.g.,>[]{liu2019quantifying}, and may help guide improvements in parameterization strategies. 

Applying these power laws to a two-year ECMWF analysis dataset in the lower tropical stratosphere reveals shallower-than-expected planetary-scale VMF cospectra of the equatorial Rossby and MRG waves, suggesting relatively excessive flux at these scales. At synoptic and subsynoptic scales, the VMF cospectra for all modes are too steep, suggesting relatively insufficient VMF at these scales. At subsynoptic scales, the majority of the VMF is associated with the IG modes, as expected. 
The data confirm the steep theoretical VMF cospectral slopes for Rossby and MRG waves at synoptic and subsynoptic scales, with the majority of their VMF confined to zonal wavenumbers below $k=10$.

The decomposition of VMFs in the two phases of the QBO provides a VMF picture consistent with theory and with the effects of the \bkg flow, highlighting the importance of the coexistance of different wave types across scales and an accurate representation of the large-scale circulation in the tropical stratosphere for the effects of both resolved and parameterized momentum fluxes. 

We showed that the cutoff scale needed to capture the majority of the Kelvin wave VMF varies with altitude in relation to the easterly and westerly shear lines, and that the first 20 zonal wavenumbers capture about 80\% of the Kelvin wave VMF flux between 5\Deg S and 5\Deg N, depending on the QBO phase and altitude. The significantly larger Kelvin wave VMF obtained in the present study, compared with earlier studies \cite<e.g.,>[]{ern2009,tindall2006equatorial,ern2009quantification}, may be related to the applied Kelvin wave filtering and absence of cutoff scale.

Within the QBO region, about 40\% of IG mode VMF is at scales with $k<20$ for both EIG and WIG modes, with differences dependent on the level and the presence of westerly and easterly shear lines. About 80\% of the IG mode VMF in the ECMWF analyses is at zonal scales smaller than 700~km ($k=60$). In the layer 100-80~hPa below the QBO region, the IG modes are the main contributors to the absolute VMF, followed by the Kelvin waves. However, their scale structure is substantially different. As a result, the majority of the zonal-momentum VMF in the lower tropical stratosphere is represented by a cutoff scale of $k=20$ to $k=30$. 
This result may help explain the smaller role of the horizontal resolution for the wave forcing of the QBO region \cite{Richteretal2014,Garfinkeletal2022,pahlavan2021revisitingPART2,polichtchouk2023resolved}, compared to the vertical resolution accompanying the gravity wave momentum flux parameterization near 100~hPa \cite{Gelleretal2016}. 

Our QBO phase comparison is limited to the region between 5\Deg S and 5\Deg N and to the two-year dataset. Future work should extend the analysis to longer datasets and  other latitudes and include the horizontal momentum fluxes as well as heat fluxes so that the full Eliasen--Palm flux vector can be obtained in the modal framework.

\section*{Data Availability Statement}
The ECMWF analysis data used in this study were produced with version CY47R2 of the Integrated Forecasting System as part of the two-year Aeolus Observing System Experiment described by \citeA{rennie2024}. Access to the original ECMWF analysis fields is subject to ECMWF data-access and licensing conditions. The processed data required to reproduce the results, figures, and tables presented in this article, together with the scripts for data analysis and visualization are published in \citeA{neduhal2026_dataset}. The new VMF decomposition method is implemented in the MODES software and will be released with the forthcoming MODES v2 software paper. 

\acknowledgments
This paper is a contribution to Subproject L2 of the Collaborative Research Centre TRR 181 “Energy Transfers in Atmosphere and Ocean”, funded by the Deutsche Forschungsgemeinschaft (DFG, German Research Foundation), project no.~274762653. We thank Frank Sielmann for contributing Figures S1--S5 in the Supplementary Information, and Frank Sielmann and Sergiy Vasylkevych for discussions during the course of this work. We also thank Branko Grisogono for reading the manuscript and providing valuable comments. 

\appendix
\section{Proportionality between VMF cospectra and horizontal and vertical kinetic energy} 
\label{theoreticalSlopeRelationshipDerivation}

We start from the equation for absolute VMF cospectra:
\begin{equation}
\label{a1}
    F_k = \sqrt{\left[\Re(\hat{u}_k\,\hat{\omega}_k^*)\right]^{2} + \left[ \Re(\hat{v}_k\,\hat{\omega}_k^*)\right]^{2}},
\end{equation}
where $\hat{u}_k$, $\hat{v}_k$, and $\hat{\omega}_k$ are complex Fourier expansion coefficients of zonal, meridional, and vertical velocities. In Euler's representation they are given by $\hat{u}_k =U_k e^{i\phi_u^{k}} $, $\hat{v}_k =V_k e^{i\phi_v^{k}} $ and $\hat{\omega}_k =W_k e^{i\phi_\omega^{k}} $ where $U_k = |\hat{u}_k| $, $ V_k =  |\hat{v}_k| $ and $W_k = |\hat{\omega}_k|$ are the moduli, and $\phi_u^{k}$, $\phi_v^{k}$ and $\phi_\omega^{k}$ are the angles of the expansion coefficients. Plugging in Eq. \ref{a1} gives us:
\begin{equation}
\label{a2}
 F_k = \sqrt{\left[\Re(U_k e^{i\phi_u^{k}} \, W_k e^{-i\phi_\omega^{k}} )\right]^{2} +\left[\Re(V_k e^{i\phi_v^{k}} \, W_k e^{-i\phi_\omega^{k}} )\right]^{2} }.
\end{equation}
$U_k$, $V_k$, and $W_k $ are positive real values and can be brought outside of the $\Re()$ operator. After rearranging, we are left with
\begin{equation}
\label{a3}
   F_k = \sqrt{U_k^{2} \, cos(\phi_u^{k}-\phi_\omega^{k})^{2} +  V_k^{2} \, cos(\phi_v^{k}-\phi_\omega^{k})^{2}  } \, |W_k|.
\end{equation}

The relative phase between $u$, $v$, and $\omega$ as a function of a zonal wavenumber can be seen in Figure~\ref{fig6} for diagonal VMFs in the tropics. For Kelvin EIG, and WIG waves relative phases do not drastically change with $k$ and they can be approximated by a constant. For MRG waves, phases appear to have one constant value on planetary and synoptic scales and another constant value on smaller scales. For Rossby waves,  phases appear to be relatively constant after the zonal wave number of $6$. Since we are considering absolute VMF cospetra independently for each diagonal term and our primary interest are slopes, we can write 
\begin{equation}
\label{a4}
  F_k= C \sqrt{U_k^{2} +  V_k^{2}  } \, |W_k|,
\end{equation}
where $C$ is some constant, or
\begin{equation}
\label{a4}
 F_k \propto \sqrt{ E_H^{k}} \sqrt{ E_V^{k}}.
\end{equation}
This approximation is less valid for zonal wavenumbers where phase spectra show rapid change.

%
%

\bibliography{litNB_AMSstyle_july2026}

%
%
%
%
%

\end{document}